\documentclass[
    aps,
    pra,
    reprint,
    superscriptaddress,
    longbibliography,
    floatfix
]{revtex4-2}

\usepackage[T1]{fontenc}
\usepackage[utf8]{inputenc}

\usepackage{amsmath, amssymb, amsfonts}
\usepackage{bm}
\usepackage{physics}
\usepackage{mathrsfs}
\usepackage{dsfont}
\usepackage[normalem]{ulem}
\usepackage{mathtools}
\usepackage{cancel}
\usepackage{ntheorem} 
\usepackage{mdframed} 

\theoremstyle{break}
\theoremheaderfont{\bfseries}
\newmdtheoremenv[%
linecolor=black,leftmargin=20,%
rightmargin=20,
backgroundcolor=gray!40,%
innertopmargin=0pt,%
ntheorem]{myprop}{Proposition}[section]

\usepackage{graphicx}

\usepackage{dcolumn}

\usepackage{xcolor}
\definecolor{dgreen}{rgb}{0,0.7,0}

\newcommand{\bluew}[1]{{\color{blue}#1}}
\definecolor{mc}{rgb}{0.9,0.3,0.2}

\usepackage{hyperref}
\hypersetup{colorlinks=true, linkcolor=blue, citecolor=blue, urlcolor=blue}

\DeclareMathOperator{\e}{\mathrm{e}}
\DeclareMathOperator{\Z}{\mathcal{Z}}

\DeclareMathOperator{\s}{\mathcal{S}}
\newcommand{\diff}{\mathop{}\!\mathrm{d}}
\newcommand{\pp}{\partial}
\newcommand{\la}{\langle}
\newcommand{\ra}{\rangle}

\newcommand{\be}{\begin{equation}}
\newcommand{\ee}{\end{equation}}

\begin{document}

    \title{Entanglement Scaling and Full Counting Statistics in Excited States
    of Two-Dimensional Rotating Fermions}

    \author{Priyangshu Goswami}
    \email{priyangshu.goswami@icts.res.in}
    \affiliation{International Centre for Theoretical Sciences, Tata Institute of Fundamental Research, Bangalore 560089, India}

    \author{Abhishek Dhar}
    \affiliation{International Centre for Theoretical Sciences, Tata Institute of Fundamental Research, Bangalore 560089, India}

    \author{Satya N. Majumdar}
    \affiliation{LPTMS, CNRS, Univ. Paris-Sud, Universit\'e Paris-Saclay - 91405 Orsay, France}

    \author{Anupam Kundu}
    \affiliation{International Centre for Theoretical Sciences, Tata Institute of Fundamental Research, Bangalore 560089, India}
    
    \date{\today}


    \begin{abstract}
        We investigate the entanglement entropy of a class of $N$-particle excited state of fermions confined in a two-dimensional harmonic trap rotating at an angular frequency $\Omega$. The excited state is constructed by filling a particular set of $N$ single-particle energy levels. We analytically compute the R\'enyi entropies of order $q$ in a disc of radius $r$ around the centre of the trap, and the cumulants corresponding to number fluctuations of fermions within the disc. We found that the area law scaling of entanglement entropy holds even for a class of excited states. We also verified the well-known series expansion of entanglement entropy in terms of the particle number cumulants for non-interacting fermions. We further derive the centered cumulant generating function, demonstrating that the associated probability distribution function in the disc has identical scaling properties, up to a variable shift, to the known ground-state result.  Finally, we extend our result to an annular region, showing that both the R\'enyi entropy and particle number cumulants of the annulus decompose into sums of the corresponding quantities of the two bounding discs. These additive relations hold as long as the width of the annulus is sufficiently large.
    \end{abstract}
    \maketitle

    \section{Introduction}
    \label{sec:intro}
    In quantum many-body systems, bipartite entanglement
    entropy provides a fundamental measure of quantum correlations between
    subsystems. Consider a system in a pure state, $\ket{\Psi}$, described by the
    density matrix $\hat\rho=\ketbra{\Psi}{\Psi}$. The R\'enyi entanglement
    entropy, $\mathcal{S}_{q}$, between a spatial domain $\mathcal{D}$ and its
    complement $\bar{\mathcal{D}}$ is defined as,
    \begin{equation}
        \s_{q}(\mathcal{D})=\frac{1}{1-q}\ln\, \mathrm{Tr}[\hat\rho_{\mathcal{D}}
        ^{q}],
    \end{equation}
    where the reduced density matrix $\hat \rho_{\mathcal{D}}$ is obtained by tracing
    out the complementary subsystem:
    $\hat\rho_{\mathcal{D}}=\mathrm{Tr}_{\mathcal{\bar{D}}}[\hat\rho]$. In the $q \to 1$
    limit, the R\'enyi entropy reduces to the von Neumann entanglement entropy ${\s_{1}}
    = -{\rm Tr}\hat\rho_{\mathcal{D}}\ln~\hat\rho_{\mathcal{D}}$.
    
    The scaling behaviour of entanglement entropy (EE) has been extensively characterized. For free fermions in
    $d$-dimensions, the ground-state EE in gapped systems,  has been shown to satisfy the area law, scaling as $\sim L^{d-1}$ \cite{RevModPhys.80.517, RevModPhys.82.277, PhysRevLett.94.060503, zeng2019gapped}.
    On the other hand, gapless non-interacting  fermionic systems exhibit a logarithmic violation of the area law scaling as $\sim L^{d-1}\ln L$ with the subsystem size $L$ \cite{Calabrese_2012, PasqualeCalabrese_2004, PhysRevB.101.235169, RevModPhys.82.277}.  Beyond uniform systems, the EE of non-interacting fermions has been investigated in the presence of external potentials \cite{PhysRevLett.112.254101, PhysRevA.91.012303}. Apart from the scaling
    properties of the EE, a connection has been established between the EE and full counting statistics (FCS) of particle number
    fluctuations, for non-interacting fermions in Gaussian
    states \cite{PhysRevA.85.062104, PhysRevLett.102.100502, song2012bipartite, 1993JETPL..58..230L,10.1063/1.531672,
    PhysRevLett.110.060602, PhysRevB.74.125315, PhysRevLett.96.076605}  (Gaussian states are defined as those where two-point correlations determine all higher order correlations, a Slater determinant being the simplest example).

    In this study, we investigate the EE and number fluctuations of non-interacting
    fermions in a two-dimensional rotating harmonic trap. This problem was first studied in \cite{PhysRevA.99.021602}
    for the ground state with a macroscopic number of particles with calculations simplified by the special structure of the the lowest Landau levels (LLL) eigenstates. The main finding was that both the EE and the second cumulant of particle number inside a central circular domain grow linearly with the perimeter (of the
    circle), establishing a direct proportionality between the two. Other
    interesting properties of this system, including the FCS and finite
    temperature effects, have been studied in \cite{PhysRevA.107.023302, PhysRevA.99.021602,
    PhysRevA.105.043315}.
    
    A natural question is as to what happens to the   entanglement area law  for excited states. Note that at finite temperatures, a natural way to specify the state of a system is using the equilibrium density matrix. However, there are no simple entanglement measures for such mixed  states and it is simpler to consider pure high energy states---for macroscopic systems one can use the usual equivalence of ensembles to associate a temperature to a pure state with specified energy.
    
    In such generic high temperature excited states, the EE is expected to scale as the volume of the subsystem~\cite{bianchi2022}. Somewhat surprisingly, it has been found that certain high-energy states can still show sub-volume entanglement. This  has been observed  in translationally invariant systems of non-interacting fermions in one and higher dimensions~\cite{Alba_2009, Angel-Ramelli_2021} and for quantum scars
    in interacting systems~\cite{Moudgalya_2022}. In the present work, we  investigate the system size scaling of EE and FCS, and the possible relationship
    between the two, for a particular class of excited states in a system of non-interacting rotating   trapped fermions. As our main result, we find that the area law continues to hold even for a class of excited states in our model.  

    For Gaussian states, the reduced density matrix is fully characterized by two-point correlations~\cite{klich2006lower, PhysRevLett.107.020601, Calabrese_2012_overlap_matrix}. Consequently, both the EE and FCS are fully determined by the eigenvalues of the correlation matrix~\cite{IngoPeschel_2003, RevModPhys.80.517}.
    For instance, the R\'enyi entropy is given in terms of the eigenvalues of the correlation matrix, $\{a_{n}\},~n=1,2,...,N$, where $0 \leq a_n \leq 1$, by
    \begin{equation}
        \s_{q}(\mathcal{D})=\frac{1}{1-q}\sum_{n=1}^{N}\, \ln\{a_{n}^{q}+
        (1-a_{n})^{q}\}, \label{eq:renyi_entropy_overlap_matrix}
    \end{equation}
    while the mean and variance of the particle number within $\mathcal{D}$ are given by,
    \begin{equation}
        \la \hat N_{\mathcal{D}}\ra=\sum_{n=1}^{N}a_{n}, \qquad \la \hat{N}^{2}_{\mathcal{D}} \ra_{c}= \sum_{n=1}^N a_n(1-a_n).\label{eq:mean.var}
    \end{equation}
    In general, the FCS provides a comprehensive characterisation
    of particle number fluctuations within the domain $\mathcal{D}$ and is widely studied 
    in contexts such as shot noise \cite{1993JETPL..58..230L}, quantum dots
    \cite{PhysRevB.74.125315}, quantum transport \cite{PhysRevLett.96.076605,10.1063/1.531672},
    spin and fermionic chains
    \cite{PhysRevE.87.022114,Abanov_2011,PhysRevLett.110.060602}, and trapped fermions
    \cite{PhysRevE.103.L030105,PhysRevLett.111.080402}.
    
    A connection between the entanglement entropy (EE) and FCS has been
    established for translationally invariant free fermionic systems~\cite{klich2006lower,
    PhysRevB.83.161408, Calabrese_2012, PhysRevLett.102.100502,
    song2012bipartite}. For Gaussian states, an exact relation exists between the
    EE and the cumulants of the particle number in $\mathcal{D}$~\cite{PhysRevB.83.161408,
    song2012bipartite}:
    \begin{align}
         & \s_{q}(\mathcal{D})=\sum_{p=1}^{\infty}\,\beta_{p}(q) \la \hat{N}^{p}_{\mathcal{D}}\ra_{c},\label{eq:renyi_cumulant_relation}
    \end{align}
    where $\hat{N}_{\mathcal{D}}$ represents the particle number operator in
    $\mathcal{D}$ and $\{\beta_{p}(q)\}$ is a set of known constants with explicit
    expressions~\cite{song2012bipartite}. This relation is also valid for non-interacting
    fermions in traps \cite{PhysRevA.85.062104, PhysRevA.107.023302}.

    Understanding entanglement properties of many-body quantum systems remains a challenging problem. For non-interacting fermions, although
    the many-body eigenstates have a simple Slater determinant structure, the computation of entanglement measures requires diagonalizing large correlation matrices. For a ground state with a large number
    of particles, this challenge has been addressed by using random matrix methods that exploit the determinantal structure of the correlations~\cite{PhysRevLett.112.254101,
    PhysRevA.91.012303, Dean_2015, PhysRevA.94.063622, Dean_2019}. A key idea used in these works was that of the overlap matrix, which allows us to compute the eigenvalues of the correlation matrix from a matrix whose entries are the single-particle inner products. This is more tractable than direct diagonalisation. In the present work we closely follow this approach.

    The paper is organized as follows. In Sec.~\ref{sec:model_methods}, we
    introduce the Hamiltonian and its single-particle spectrum, and describe the
    class of excited states under consideration. We also provide a summary of
    the main results and of the methods used. In Sec.~\ref{section:disc_entropy},
    we evaluate the entanglement entropy (EE) using the overlap matrix obtained previously
    and analyze its scaling behavior. Then, we use these eigenvalues to determine
    the full counting statistics: we study the generating function of particle-number
    cumulants, compute the cumulants and their scaling behavior, and verify the
    cumulant expansion of the EE. In Sec.~\ref{sec:annulus}, we compute the EE for
    an annular region and demonstrate its additivity in terms of the corresponding
    disc contributions. Finally, Sec.~\ref{section:conclusion} summarizes our
    results and outlines possible future directions.

    \section{Model, Summary of results and Methods}
    \label{sec:model_methods}

    \subsection{Definition of model and states}
    Let us consider $N$ non-interacting spinless fermions confined in a trap in two
    dimensions. Let $\phi_{i}(\vec x)$ denote the $i$-th single particle
    eigenfunction with energy $\epsilon_{i}$. Any many body fermionic eigenstate
    can be constructed from a set of $N$ single particle eigenfunctions in the form
    of an $(N\times N)$ Slater determinant
    \begin{equation}
        \Psi_{E}\left(\vec x_{1},\vec x_{2},\ldots, \vec x_{N}\right)= \frac{1}{\sqrt{N!}}
        \,{\rm det}\left[\phi_{i}(\vec x_{j})\right]\, , \label{slater.1}
    \end{equation}
    where $i$ and $j$ run over the occupied orbitals and particle coordinates,
    respectively. The eigenenergy of this many--body state is
    \begin{equation}
        E= \sum_{i=1}^{N}\epsilon_{i}\, , \label{energy.1}
    \end{equation}
    with $\epsilon_{i}$ the single-particle energies of the eigenstates in the
    many body state. Choosing the lowest $N$ levels yields the ground state $\Psi
    _{E_0}(\{\vec x_{i}\})$, whose energy is the sum of the first $N$ single-particle
    eigenvalues. Any other choice corresponds to an excited state with energy given
    by Eq.~\eqref{energy.1}.

    Specifically we consider a two-dimensional harmonic trap characterised by a frequency
    $\omega$, and rotating at an angular velocity $\Omega$. Such systems are of significant
    interest in both theoretical \cite{PhysRevLett.85.4648, PhysRevLett.87.060403,
    PhysRevA.71.023611, tonini2006formation,PhysRevLett.90.140402} and experimental \cite{PhysRevLett.92.040404,
    zwierlein} studies, particularly within the context of cold atom systems
    \cite{cooper2008rapidly}. Notably, in the regime where
    $\Omega \lesssim \omega$, this system exhibits an analogy to quantum Hall systems
    \cite{RevModPhys.80.885}.

    In the reference frame rotating with angular velocity $\Omega$, the many-body
    Hamiltonian $\hat{\mathcal{H}}$ of $N$ non-interacting spinless fermions
    confined in a two-dimensional harmonic trap with frequency $\omega$ reads
    \begin{align}
        \hat{\mathcal{H}}= & \sum_{j=1}^{N}\left[ \frac{\Hat{{\bf p}_j}^{2}}{2m}+\frac{1}{2}m\omega^{2}\hat{{\bf r}}_{j}^{2}-\Omega \Hat{L}^{z}_{j}\right]\, ,\label{eq:single_p_ham}
    \end{align}
    where $\hat{{\bf p}}_{j}$ and $\hat{{\bf r}}_{j}$ are the momentum and
    position operators for the $j$-th particle and
    $\Hat{L}^{z}_{j}=i \hbar (\hat y_{j}\pp_{x_j}-\hat x_{j}\pp_{y_j})$ is the
    angular momentum in the $z$-direction. To keep the fermions trapped, we
    assume $\omega > \Omega$. The single particle spectrum of this system has been
    derived in \cite{PhysRevLett.85.4648}, with the wavefunction and energies
    given by
    \begin{align}
        \phi_{k,l}(x,y) & =\frac{e^{|z|^2/2}\partial_{+}^{k}\partial_{-}^{l}e^{-|z|^2}}{\sqrt{\pi a^{2}~ k!~ l!}},                                              \\
        \epsilon_{k,l}  & =\hbar \omega\left\{ 1+k\left(1-\frac{\Omega}{\omega}\right)+l\left(1+\frac{\Omega}{\omega}\right)\right\},\label{eq:energy_spectrum}
    \end{align}
    where $k,l=0,1,2...$, $z=x+iy$,
    $\partial_{\pm}=(\partial_{x}\pm i\partial_{y})/2$ and we are measuring
    lengths in units of the scale, $a=\sqrt{\hbar/m \omega}$. Here $l$ denotes the Landau levels, with $l=0$ marking the lowest Landau level (LLL). It can be shown
    that $(k-l)$ is in fact the angular momentum index. Henceforth, without loss of
    generality, we will set $\omega=1$ and measure energy in units of $\hbar \omega$.
    Note that we need $\Omega <1$ for the spectrum to be bounded.

    We will focus on the particular set of single particle states in the LLL with $l=0$ for
    which the single-particle wavefunctions (in polar coordinates) and energy
    levels are, from Eq.~\eqref{eq:energy_spectrum}, given by,
    \begin{align}
        \phi_{k}(r,\theta) \coloneqq \phi_{k,0} & =\frac{(-1)^{k}r^{k}e^{-r^2/2}e^{ik \theta}}{\sqrt{\pi k!}},  \label{eigenfunction}  \\
        \epsilon_{k}\coloneqq \epsilon_{k,0}    & = 1+k(1-\Omega),\quad k=0,1,2,\ldots. \label{energy_spectrum-1}
    \end{align}
    Then, we construct an $N$-particle many-body state by filling levels $k=M$ to $k=M+N-1$. For $M \geq 0$, this, in general,  yields an excited state.

    This sliding window construction was first studied analytically in a one-dimensional harmonic trap in~\cite{Cunden_2019}, where, the properties of a many-body excited state, consisting of $N$ consecutive single particle levels with energies from $(M+1/2)$ to $(M+N-1/2)$ (in units of $\hbar \omega$) were computed exactly for arbitrary $M$ and $N$. In particular, the average density and two-point correlation kernel were obtained in closed form, and tuning $M$ was shown to interpolate continuously through a family of excited many-body states.
    
    Here, we extend this idea to non-interacting fermions in a two-dimensional rotating harmonic trap. In \cite{PhysRevA.99.021602}, the rotation frequency was constrained to $1 \, > \Omega >\left(1-\frac{2}{M+N}\right)$, ensuring that the $N$ chosen levels are consecutive in energy and the state fully belongs to the lowest Landau level (LLL) sector. In the present work, we retain the sliding window structure-$N$ consecutive $k$ values at $l=0$-but impose no such constraint on $\Omega$. Consequently, the selected levels are consecutive in $k$ but may not be in energy. The state is a genuine excited state of the 2D harmonic trap for $M>0$.
    
\begin{figure}[ht!]
        \centering
        \includegraphics[width=\columnwidth]{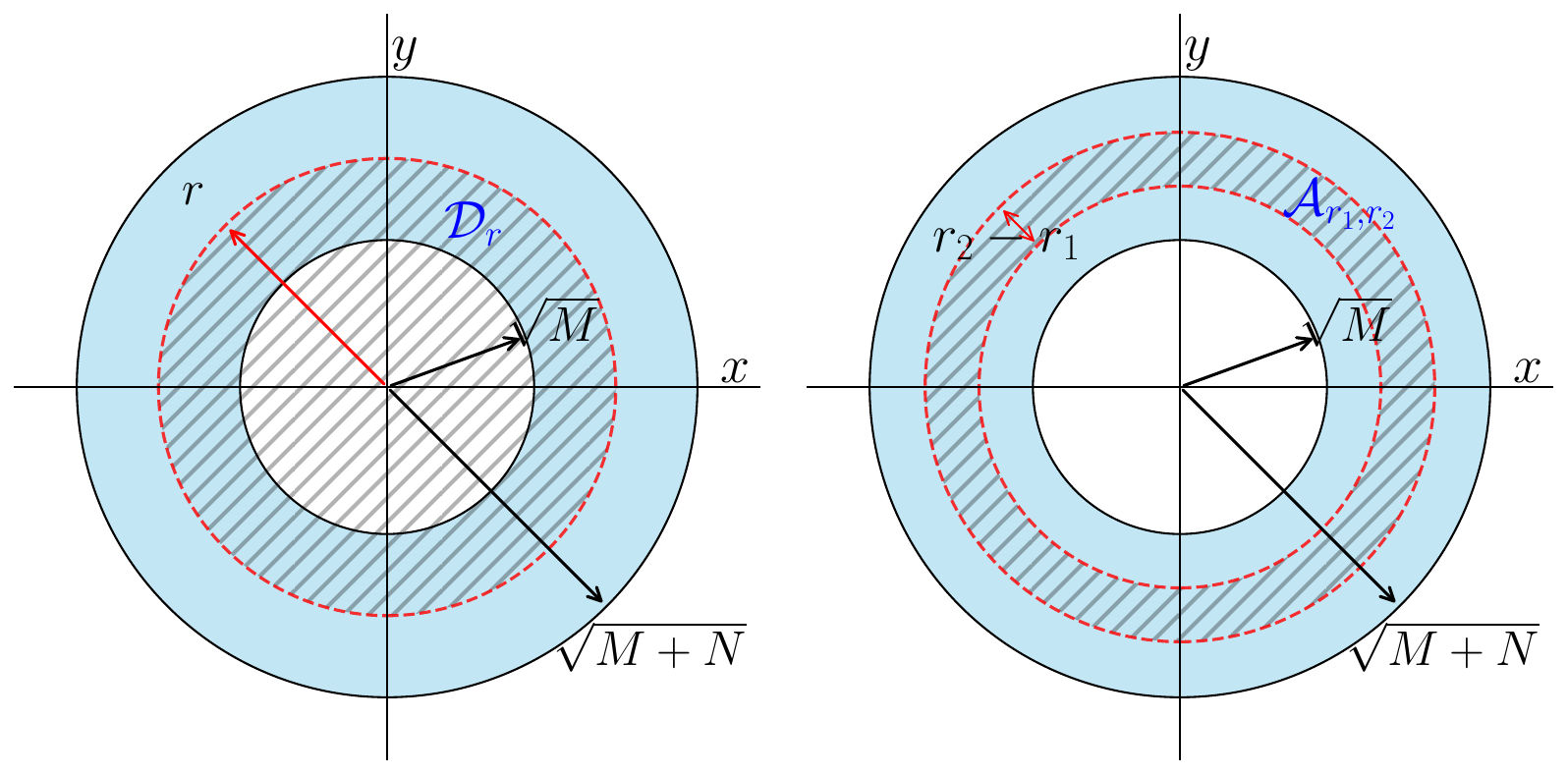}
        \caption{For the excited state $ \ket{\Psi_{(M,M+N)}} $, the fermions are mostly confined uniformly in the blue coloured region where $r\in \left(\sqrt{M},\sqrt{M+N}\right)$. The partition domains used for computing EE and FCS are indicated by the dashed regions-- (left) disc $\mathcal{D}_{r}$ and (right) annulus $\mathcal{A}_{r_{1},r_{2}}$.}
        \label{fig:particle_density_profile}
    \end{figure}

    Let us denote our many-body state by $\ket{\Psi_{(M,M+N)}}$, with energy $E=\sum_{k=M}^{N-1+M}\epsilon_{k}$ and a corresponding density matrix, $\hat\rho=\ketbra{\Psi_{(M,M+N)}}$. Our main interest is in computing the EE and FCS of particles within a radially symmetric domain: either a disc $\mathcal{D}_{r}$ of radius $r$, or an annulus $\mathcal{A}_{r_1,r_2}$ with inner and outer radii $r_1$ and $r_2$. The structure of the $l=0$ single-particle states makes these observables exactly computable and allows us to extend the results of \cite{PhysRevA.99.021602} to arbitrary $M$. The simplifying feature is to use the overlap matrix (Sec.~\ref{subsection:overlap_matrix}) whose diagonal structure reduces the calculation of the EE and FCS to spectral sums over eigenvalues. In Fig.~\ref{fig:particle_density_profile}, we show the domains $\mathcal{D}_{r}$ and $\mathcal{A}_{r_{1},r_{2}}$ overlaid on the particle density profile of the excited state.

    \subsection{Summary of results}
    We briefly summarise our results as follows.
    \begin{enumerate}
        \item In the excited many-body state $\ket{\Psi_{(M, M+N)}}$ and in the limit of large $M$ and large $N$, with the ratio $\frac{M}{N}=\mu$ fixed, fermions are restricted in a ring-shaped region between the radii $\sqrt{M}$ and $\sqrt{M+N}$, where the average density becomes uniform.

        \item Within the disc region $\mathcal{D}_{r}$, both the EE and all even cumulants of $\hat{N}_{\mathcal{D}_r}$ increase proportionally with the radius, $r$, provided one is in the \textit{bulk} region ($\sqrt{M}\ll r \ll \sqrt{M+N}$). 

        \item The EE and cumulants at the \textit{inner} edge ($|r-\sqrt{M}| \sim
            O(1)$) and \textit{outer} edge ($|r-\sqrt{M+N}| \sim O(1)$) demonstrate
            a universal scaling behavior as $M$ and $N$ becomes large, keeping $\mu$ fixed. These scaling functions are determined
            solely by the distance from the respective edges.

        \item  We verified the cumulant expansion of the EE, given in Eq.~\eqref{eq:renyi_cumulant_relation}, for the domain $\mathcal{D}_{r}$. We observed a fast convergence of the series for $q\ge 2$.
            
        \item In the limit of large $M$ and $N$ (keeping the ratio fixed), the centred cumulant generating
            function (CCGF) for the disc region in the excited state reveals
            that the probability distribution function of $\hat{N}_{\mathcal{D}_r}$ has the same scaling properties, up to variable shift by $\mu=\frac{M}{N}$, as that of the $M=0$ state studied previously
            in \cite{allez2014index, shirai2006large, PhysRevE.100.012137}, .

        \item For an annulus $\mathcal{A}_{r_1,r_2}$ located well
            inside the \textit{bulk} region, we found that the EE can be broken down into the sum
            of the EEs of the discs $\mathcal{D}_{r_{1}}$ and $\mathcal{D}_{r_{2}}$
            if $|r_{2}-r_{1}| > O(1)$. We demonstrate that this additive
            property is also applicable to particle number cumulants, as long as
            the boundaries of the annulus do not interact with each other.
    \end{enumerate}

    \subsection{Methods: overlap Matrix, its eigenvalues and density profiles}
    \label{subsection:overlap_matrix} We first note that for any pure Gaussian state
    of non-interacting Fermions, both the EE and FCS can be expressed in terms
    of the eigenvalues of the correlation matrix defined over the domain of
    interest. For the state formed with $N$ occupied particle states,
    $\phi_{k}({\bf r})$ with corresponding energies $\epsilon_{k}$, the correlation kernel is defined as
    \begin{align}
        C({\bf x},{\bf x'}) = \sum_{k}\phi_{k}^{*}({\bf x}) \phi_{k}({\bf x'}).
    \end{align}
    The eigenvalues of this correlation kernel are defined through the integral equation
    $\int_{A}C({\bf x},{\bf y}) f_{s}({\bf y}) \diff{{\bf y}}= a_{s}\, f_{s}({\bf x}
    )$, with $\mathscr{A}$ indicating the spatial domain for which we want to compute the
    entanglement. It can be shown that there are exactly $N$ non-vanishing eigenvalues
    which can equivalently be obtained by diagonalizing the $N\times N$ overlap matrix defined
    via
    \begin{align}
        \mathbb{A}_{kl}= \int_{A}\diff{\bf x}~\phi_{k}^{*}({\bf x}) \phi_{l}({\bf x}).
    \end{align}
    In terms of the $N$ eigenvalues $a_{s},~s=1,2\ldots,N$ of the overlap matrix,
    the Renyi EE is given by Eq.~\eqref{eq:renyi_entropy_overlap_matrix}. For the
    FCS one often defines the moment generating function for the number of
    fermions, $\hat N_{A}$, inside the domain $A$, ${\cal Z}(\lambda;A)= \la e^{-\lambda {\hat N}_A}\ra$, where $\la ...\ra$ denotes expectation value in the pure state. It can also be expressed in terms of the overlap matrix eigenvalues as:
    \begin{align}
        \ln \mathcal{Z}(\lambda;A) =\sum_{s=1}^{N}\ln\!\left[1+(e^{-\lambda}-1)a_{s}\right]. \label{eq:cumgf}
    \end{align}
    It was shown in
    \cite{PhysRevLett.107.020601, Calabrese_2012_overlap_matrix, PhysRevA.91.012303}
    that the overlap-matrix formalism provides an efficient framework for
    computing both the entanglement entropy (EE) and full counting statistics (FCS).
    In particular, it was demonstrated in \cite{PhysRevA.99.021602} that, this method
    greatly simplifies calculations for the $M=0$ state. In the present work, we
    adopt this formalism because a similar simplification naturally emerges for the excited
    states under consideration.

    Let us recall that our $N$-particle state is constructed using $N$ levels $k=
    M,M+1,\ldots,M+N-1$ from the set of single particle states in Eq.~\eqref{energy_spectrum-1}.
    For this set and the choice of radially symmetric domains $\mathscr{A}$, the
    simplfication follows from the fact that the overlap matrix is diagonal and
    hence the eigenvalues are known explicitly.

    The overlap matrix $\mathbb{A}$ associated with a given spatial region is
    defined as the restriction of the single-particle inner product to that
    region. For the disc region $\mathcal{D}_{r}$, the matrix elements are
    \begin{equation}
        \mathbb{A}_{n_{1},n_{2}}=\int_{\mathcal{D}_{r}}\, \diff{r}\diff{\theta}\,
        r\, \phi_{n_{1}}^{*}(r,\theta)\, \phi_{n_{2}}(r,\theta),
    \end{equation}
    where $\phi_k^*(r, \theta)$ is given in Eq.~\eqref{eigenfunction}. Also, $n_{1}$ and $n_{2}$ run through the filled energy levels $M$ to $M+N-1$. One immediately gets the diagonal form
    \begin{align}
        \begin{split}\mathbb{A}_{n_{1},n_{2}}=&a_{n_{1}+1-M}(r)\,\delta_{n_{1},n_{2}}, \\{\rm where}~~ a_{s}(r)=&\frac{\gamma(s+M,r^{2})}{(s+M-1)!}\end{split} \label{eq:disc_eigenvalues},
    \end{align}
    and $s=1,2,\ldots,N$, and $\gamma(k,r^{2})$ is the incomplete gamma function
    defined as
    \begin{equation}
        \gamma(k,r^{2})=\int_{0}^{r^{2}}\diff{x}\,\e^{-x}\,x^{k-1}. \label{eq:incomplete_gamma}
    \end{equation}
    Thus, the overlap matrix in the disc region is diagonal, and we have
    explicit forms for the eigenvalues $a_{s}(r)$, $s=1,2\ldots,N$.

    For the case of the annular region $\mathcal{A}_{r_{1},r_{2}}: r_{1}<|z|<r_{2}$,
    a similar computation gives:
    \begin{align}
        \label{eq:annulus_eigenvalues_0}\begin{split}\tilde{\mathbb{A}}_{(n_{1},n_{2})}=&\tilde{a}_{n_{1}+1-M}(r_{1},r_{2})\,\delta_{n_{1},n_{2}}, \\{\rm with}~~ \Tilde{a}_{s}(r_{1},r_{2})=&a_{s}(r_{2})-a_{s}(r_{1}).\end{split}
    \end{align}
    This provides a simple and intuitive geometric interpretation: the annulus
    contribution is obtained as the difference between two concentric discs. The
    eigenvalues from  Eqs.~\eqref{eq:disc_eigenvalues} and~\eqref{eq:annulus_eigenvalues_0} can
    be used to evaluate the Rényi entanglement entropy using Eq.~\eqref{eq:renyi_entropy_overlap_matrix}.
    Moreover, they also allow us to compute the particle-number cumulants in both
    the disc $\mathcal{D}_{r}$ and the annulus $\mathcal{A}_{r_{1}, r_{2}}$,
    using Eq.~\eqref{eq:cumgf}. We provide details of these computations in Sec.~\ref{section:disc_entropy}
    for the disc region and in Sec.~\ref{sec:annulus} for the annular region.

    \subsection{Particle Number Density Profile}
    \label{subsection:particle_density}

    In this section, we discuss the density profiles corresponding to the excited state. This density is rotationally symmetric and depends only on the radial coordinate $r$, and is given by
    \begin{align}
        \rho(r) = \sum_{k=M}^{M+N-1}|\phi_{k}(r,\theta)|^{2}
                = \sum_{k=M}^{M+N-1}\frac{r^{2k}e^{-r^{2}}}{\pi k!}.
    \label{eq:density_profile}
    \end{align}
    We assume that $M, N$ are large, keeping $\mu=\frac{M}{N}$ fixed. Because consecutive terms vary slowly with $k$, in the large $N$ limit we can convert the sum to an integral,
    \begin{equation}
        \rho(r) \approx \int_{M}^{M+N}\frac{\diff{k}}{\pi}\,\frac{r^{2k}e^{-r^{2}}}{k!}.
    \end{equation}
    Using Stirling's approximation for $k!$ and introducing the scaling variables $\xi=\frac{r}{\sqrt{N}},~~ v^{2}=\frac{k}{N}$, we obtain,
    \begin{equation}
        \rho(r)\approx \frac{2}{\pi}\sqrt{\frac{N}{2\pi}}
        \int_{\sqrt{\mu}}^{\sqrt{\mu+1}}\diff{v}\,e^{Nf_{\xi}(v)},\label{eq:saddle.point}
    \end{equation}
    where
    \begin{equation}
        f_{\xi}(v)=2v^{2}\ln\xi-\xi^{2}-2v^{2}\ln v+v^{2}.
    \end{equation}
    The integral in Eq.~\eqref{eq:saddle.point} can be performed using saddle point method. We note that the saddle point is at $v^{*}=\xi$, where the exponent vanishes, $f_{\xi}(v^*)=0$. Expanding the exponent around $v^*$ to quadratic order, we get,
    \begin{equation}
        \rho(r)\approx \frac{2}{\pi}\sqrt{\frac{N}{2\pi}}
        \int_{\sqrt{\mu}}^{\sqrt{\mu+1}}\diff{v}\,e^{-2N(\xi-v)^{2}},
        \label{eq:approx_density_profile}
    \end{equation}
    which on performing the integral gives,
    \begin{align}
        \rho(r=\sqrt{N}\xi)\approx & \frac{1}{2\pi}\,\mathrm{erfc}\!\left[\sqrt{2N}\left(\xi-\sqrt{\mu+1}\right)\right]\nonumber \\
        -              & \frac{1}{2\pi}\,\mathrm{erfc}\!\left[\sqrt{2N}\left(\xi-\sqrt{\mu}\right)\right], \label{eq:approx_density_profile1}
    \end{align}
    where $\mathrm{erfc}(x)=\left(\frac{2}{\sqrt{\pi}}\right)\int_{x}^{\infty}\diff{z}\,e^{-z^{2}}$.
    
\begin{figure}[htb!]
        \centering
        \includegraphics[width=\columnwidth]{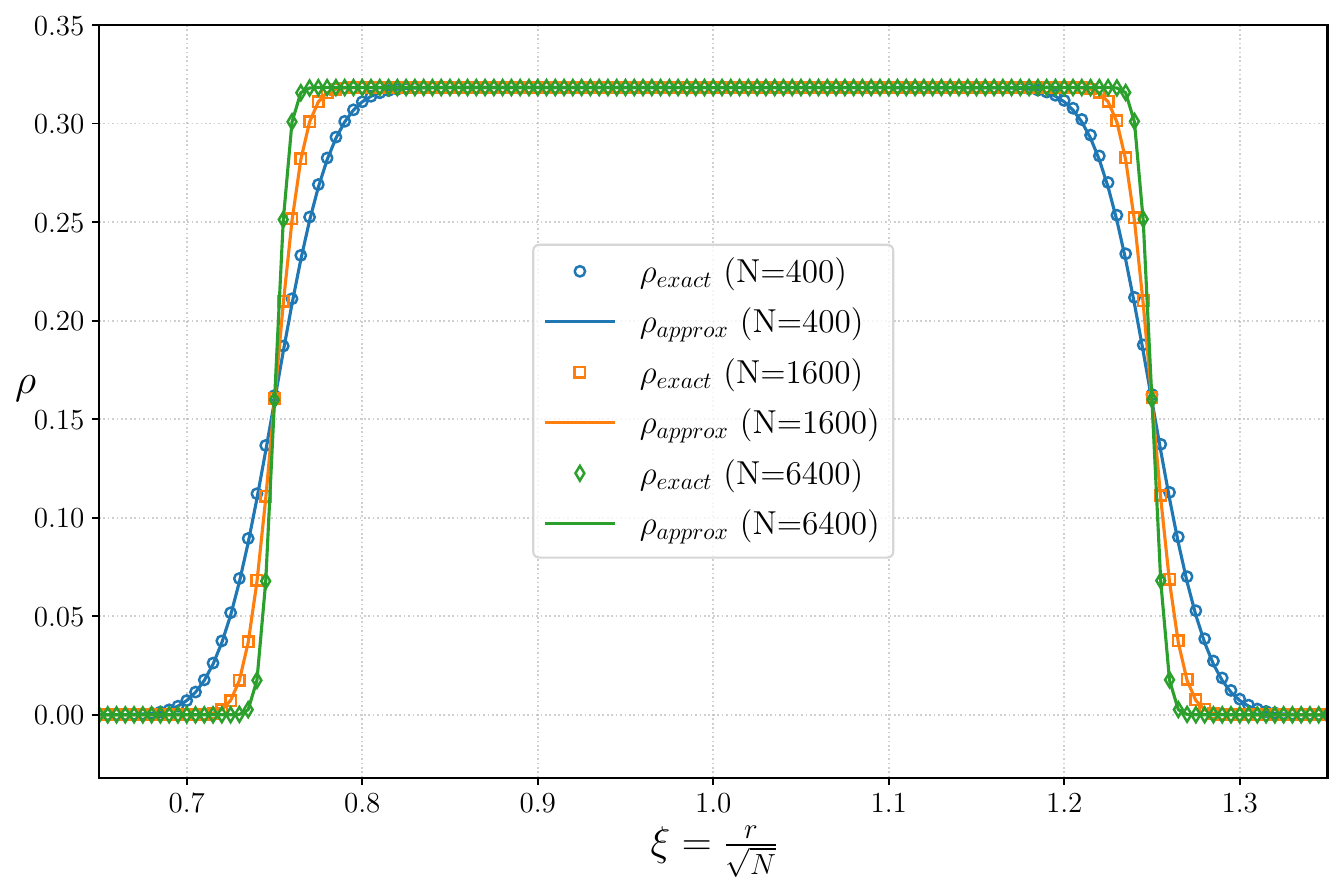}
        \caption{
        Radial density profile $\rho(r)$ for excited states with $\mu = M/N = 9/16$ and $N=400,1600,6400$. Scatter points represent the exact expression in Eq.~\eqref{eq:density_profile} while the solid lines represent the approximation in Eq.~\eqref{eq:approx_density_profile1}. The approximation reproduces the uniform bulk density and the edge behaviour quite well.
        }
        \label{fig:density_profile}
    \end{figure}

    In Fig.~\ref{fig:density_profile}, we compare the exact radial density
    profile $\rho(r)$ from Eq.~\eqref{eq:density_profile} with the approximate expression
    in Eq.~\eqref{eq:approx_density_profile1} for excited states characterised by
    a fixed $\mu=9/16$. The approximation agrees well with the exact result. We observe
    that the density profile has three regimes --- the \textit{inner edge} ($|\xi
    -\sqrt{\mu}| \simeq N^{-1/2})$, \textit{bulk} ($\sqrt{\mu} < \xi < \sqrt{1+\mu}
    )$, and \textit{outer edge} ($|\xi-\sqrt{1+\mu}| \simeq N^{-1/2})$. In the $N\to \infty$ limit, using ${\rm erfc}(-\infty)=2$ and ${\rm erfc}(\infty)=0$, Eq.~\eqref{eq:approx_density_profile1}
    predicts value of the density equal to $\frac{1}{\pi}$ in the bulk region. The mean of the total
    number of particles in the domain $\mathcal{D}_{r}$, \textit{i.e.,} the mean of $\hat{N}_{\mathcal{D}_{r}}\equiv \hat{N}_{r}$, can be obtained by integrating the density profile as,
    \begin{subequations}
    \label{def:<N_r>-tot}
    \begin{equation}
        \label{def:<N_r>}\la \hat{N}_{r}\ra=\sum_{s=1}^{N}a_{s}(r)\approx 2 \pi \int_{0}^{r}\diff{r'}\, r'\rho
        (r').
    \end{equation}
    Using the approximate form of $\rho(r)$ in  Eq.~\eqref{eq:approx_density_profile1}, the above integral yields the following asymptotic behaviour in the limit of large $N$
    \begin{align}
        \la \hat{N}_{r}\ra\approx N~\mathcal{F}\left( \frac{r}{\sqrt{N}}\right), \label{<N>-scaling}
    \end{align}
    where $\mathcal{F}(\xi)=\left( \xi^{2}-\mu\right)$ for
    $\sqrt{\mu}< \xi< \sqrt{1+\mu}$, zero for $\xi \le \sqrt{\mu}$ and one for $\xi \geq \sqrt{1+\mu}$.
\end{subequations}

    \section{Entanglement and FCS of Disc Region}
    \label{section:disc_entropy}
    \subsection{Entanglement}
    \label{subsec:disc_entanglement}
    Given the explicit overlap matrix eigenvalues, Eq.~\eqref{eq:disc_eigenvalues}, we use the expression in Eq.~\eqref{eq:renyi_entropy_overlap_matrix} to obtain the $q$-th R\'enyi entropy,  $\s_{(q)}(\mathcal{D}_{r})\equiv \s_{q}(M,N;r)$, for the disc $\mathcal{D}_{r}$:
    \begin{equation}
    \s_{q}(M,N;r)=\frac{1}{1-q}\sum_{s=1}^{N}\ln\left[a_{s}(r)^{q}+(1-a_{s}(r))^{q}\right].\label{eq:disc_s_q.1}
    \end{equation}
    The set of eigenvalues $\{a_{s}(r)\}$ can be represented via the integral
    form of the incomplete gamma function in Eq.~\eqref{eq:incomplete_gamma}. We recall that we are working in the limit of large $M$ and $N$, keeping $\frac{M}{N}=\mu$ fixed and $r \sim O(\sqrt{N})$. In this limit, a saddle point approximation of Eq.~\eqref{eq:incomplete_gamma} yields the form \cite{DLMF:8.11}:
    \begin{equation}
    \label{eq:asymptotic_eigenvalue}
    \begin{split}
    a_{s}(r)&\approx \tfrac{1}{2}\,\mathrm{erfc}(y(s;r)),\\
    \text{where}\quad y(s;r)&= \frac{M+s-1-r^{2}}{\sqrt{2(M+s-1)}},
    \end{split}
    \end{equation}
    and the identity $\mathrm{erfc}(y)+\mathrm{erfc}(-y)=2$ gives directly $1-a_{s}(r)\approx \tfrac{1}{2}\,\mathrm{erfc}(-y(s;r))$. We now make a change of variable from $s$ to $z=\frac{M+s-r^{2}}{2r}$ and note that $\diff{z}=\frac{1}{2r}$ is small since $r\sim O(\sqrt{N})$. Hence, we can convert the sum in Eq.~\eqref{eq:disc_s_q.1} to an integral, and get,
    \begin{equation}
    \label{eq:exact_integral_0}
    \begin{split}
    &\s_{q}(\mu N,N;r)\approx~(2\pi r)\times\mathcal{I}_{q}(\mu,N; r),~~~~{\rm where,}\\
    &\mathcal{I}_{q}(\mu,N; r)=\,\frac{1}{\pi(1-q)} \int_{\frac{\mu N-r^{2}}{2r}}^{\frac{(\mu+1)N-r^{2}}{2r}}\diff{z}\\
    &~~~~~\ln\left[\left\{\tfrac{1}{2}\mathrm{erfc}\left(z\sqrt{2}
    \left(1+\tfrac{2z}{r}\right)^{-1/2}\right)\right\}^{q}+\right.\\
    &~~~\qquad\left.\left\{\tfrac{1}{2}\mathrm{erfc}\!\left(-z\sqrt{2}
    \left(1+\tfrac{2z}{r}\right)^{-1/2}\right)\right\}^{q}\right].
    \end{split}
    \end{equation}
    
    As shown in Appendix~\ref{appendix:ee_for_disc}, the function $\mathcal{I}_{q}
    (\mu,N; r)$, for large $N$, has the following scaling forms in different regimes,
    \begin{align}
         & \mathcal{I}_{q}(\mu,N; \sqrt{N}\xi )\label{eq:i_q.1}                                                                                                                                                                                                                                                        \\
         & ~~\approx\begin{cases}F_{q}\left(\sqrt{N}\left(\xi-\sqrt{\mu}\right) \right),~&~|\xi-\sqrt{\mu}| \lesssim \frac{1}{\sqrt{N}}\\ \sigma_{q},~&~\sqrt{\mu}\ll \xi \ll \sqrt{1+\mu}\\ F_{q}\left(-\sqrt{N}\left(\xi-\sqrt{1+\mu}\right) \right),~&~|\xi-\sqrt{1+\mu}| \lesssim \frac{1}{\sqrt{N}}\\\end{cases} \notag
    \end{align}
    where
    \begin{align}
        F_{q} & (y)=[\sqrt{2}\pi(1-q)]^{-1}\int_{-\infty}^{y\sqrt{2}}\,\diff{z}\nonumber                                                                                                                         \\
              & ~~~~~\ln\left[\left\{\frac{1}{2}\mathrm{erfc}\left(z\right)\right\}^{q}+\left\{\frac{1}{2}\mathrm{erfc}\left(-z\right)\right\}^{q}\right], \label{eq:ee_scaling_function}
    \end{align}
    and
    \begin{align}
        \sigma_{q}:=F_{q}(\infty).\label{eq:sigma_q}
    \end{align}
    Thus, for large $N$, we have following scaling form for the Rényi entropy:
    \begin{align}
         & \frac{\s_{q}(\mu,N;\sqrt{N}\xi)}{\sqrt{N}}\approx 2\pi \xi \times \mathcal{I}_{q}(\mu,N; \sqrt{N}\xi ), \label{eq:mcS_q-scaling}
    \end{align}
    where $\mathcal{I}_{q}$ is defined in Eq.~\eqref{eq:i_q.1}.
    \begin{figure}[htb!]
        \centering
        \includegraphics[width=\columnwidth]{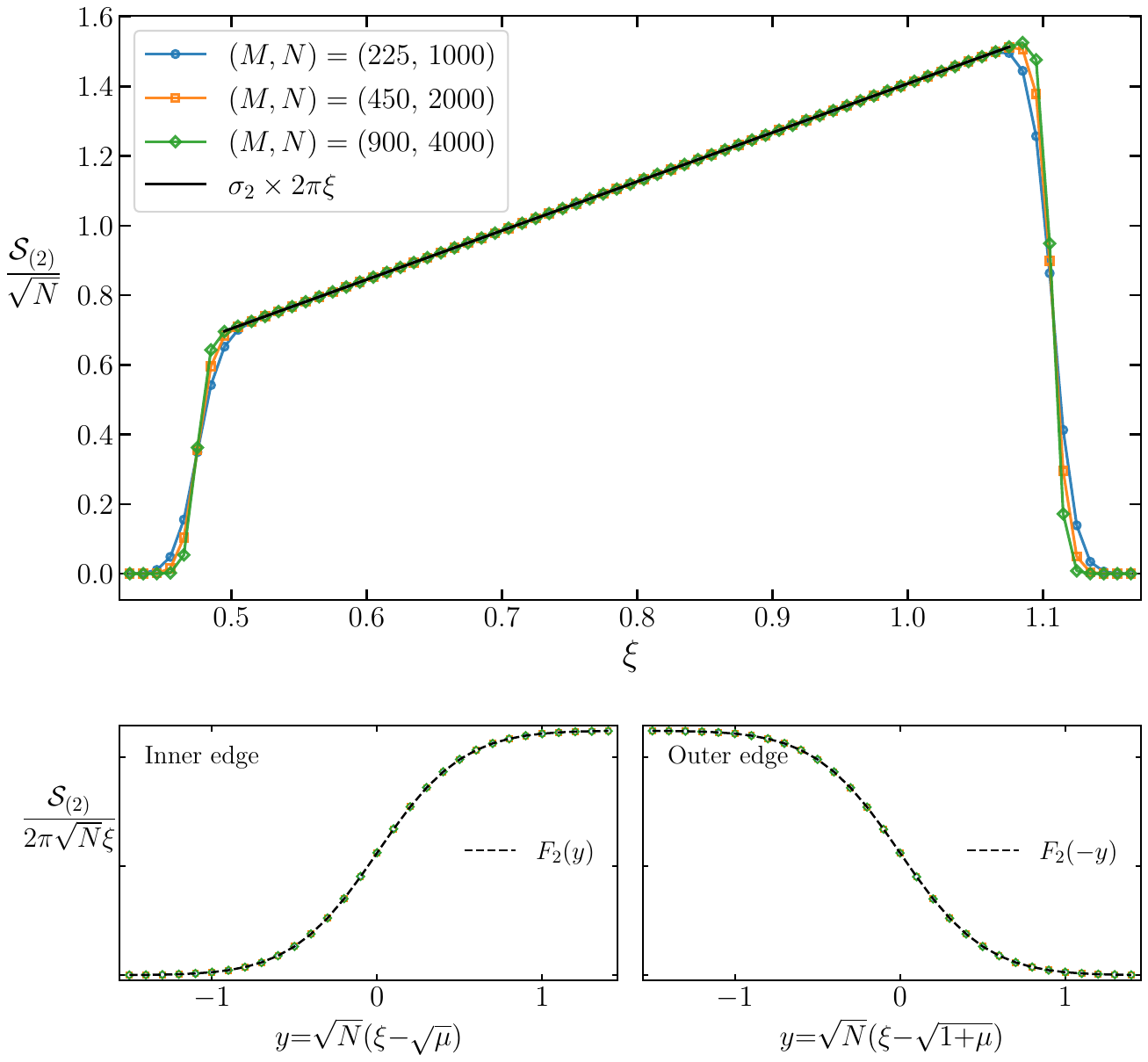}
        \caption{The upper panel shows the second R\'enyi entropy $\s_{(2)}$ as a function of the rescaled radial coordinate $\xi$, for several values of $(M,N)$ at fixed $\mu=0.225$. Linear growth in the \textit{bulk} region is confirmed by a straight-line fit, with area-law coefficient $\sigma_{2} = F_{2}(\infty)$ from Eq.~\eqref{eq:ee_scaling_function}. The lower panel compares the numerically computed ratio $\s_{(2)}/(2\pi\sqrt{N}\xi)$ (marked points) with the theoretical scaling function $F_{2}$ from Eq.~\eqref{eq:i_q.1} (dashed lines) at the inner and outer edge regions.}
        \label{fig:renyi_zeta}
    \end{figure}
    Figure~\ref{fig:renyi_zeta} shows the agreement of the scaling behaviour from
    Eqs.~\eqref{eq:i_q.1} and~\eqref{eq:mcS_q-scaling} with the numerically generated
    data for $\mu=0.225$ and $N=1000,2000,4000$. In the bottom panel of \ref{fig:renyi_zeta}, we have
    also matched the behaviour of the function in Eq.~\eqref{eq:i_q.1} at the edge regions with the numerically generated
    ratio of the R\'enyi entropy to the disc circumference. The graphical evidence
    clearly confirms that, although the bulk region adheres to an area law
    scaling, the edges are influenced by crossover functions dictated by the complementary
    error function structure of the underlying overlap-matrix eigenvalues.
    Setting $\mu=0$, we can reproduce the R\'enyi entropy for the $\ket{\Psi_{(0,N)}}$ state,
    whose behaviour has been studied in \cite{PhysRevA.99.021602}. The area-law scaling
    coefficient $\sigma_{q}$ in this work matches with the value obtained  for $\ket{\Psi_{(0,N)}}$ state previously.

    This area law scaling behaviour is typically anticipated for states where the
    2-point correlation function decreases exponentially as a function of separation \cite{brandao2013,
    Calabrese_2012_overlap_matrix, RevModPhys.82.277}.
    
    %

    \subsection{Full Counting Statistics}
    \label{subsec:fcs_disc}
    \noindent
    To characterise number fluctuations around the mean, we require the centred
    cumulant generating function (CCGF),
    \begin{equation}
        \chi(\lambda; r) = \ln \la e^{-\lambda(\hat N_{r}-\la \hat N_{r} \ra)} \ra =\ln \mathcal{Z}(\lambda; r)+\lambda \la \hat{N}_{r}\ra , \label{ccgf_logarithm}
    \end{equation}
    where $\mathcal{Z}(\lambda; r)\equiv {\cal Z}(\lambda; {\cal D}_{r})$ is defined in Eq.~\eqref{eq:cumgf} and $\la \hat
    {N}_{r}\ra = \sum_{s=1}^{N}a_{s}(r)$ is given in Eq.~\eqref{def:<N_r>-tot}. From the discussion of
    the density in Sec.~\ref{subsection:particle_density}, we observe that the
    particles, on an average, are accumulated in the intermediate regime
    $\sqrt{\mu}\le \xi \le \sqrt{1+\mu}$ corresponding to the bulk of the
    fermionic cloud. In this region, we expect non-trivial fluctuations which we
    analyse below.

    A Taylor expansion of the CCGF $\chi(\lambda;r)$ with respect to $\lambda$
    gives the higher cumulants, i.e.,
    \begin{equation}
        \chi(\lambda; r)= \sum_{p=2}^{\infty}  \frac{(-\lambda)^p}{p!}\la \delta \hat{N}^{p}_{r}\ra_{c}, 
    \end{equation}
    where $\la \delta \hat{N}^{p}_{r}\ra_{c}$ is the $p$-th cumulant of the centered variable $ \delta \hat{N}_r=\hat N_r-\la \hat N_r\ra$. Note that, $\la \delta \hat{N}^{p}_{r}\ra_{c}=\la  \hat{N}^{p}_{r}\ra_{c}$ for all $p\ge 2$. A calculation analogous to what is done in \cite{PhysRevA.99.021602} for the case of the $M=0$ state is performed in Appendix \ref{app:ccgf_disc_appe} and we find the cumulants for $p \ge 2$ (note that the central cumulant is zero by definition for $p=1$) as functions of the overlap-matrix eigenvalues,
    \begin{align}
        \begin{split}\la \hat{N}^{p}_{r}\ra_{c}&=\sum_{s=1}^{N}\, n_{p}(a_{s}(r)),\quad {\rm where,}\\
        n_{p}(x)&=\begin{cases}-\mathrm{Li}_{1-p}\left(-\frac{x}{1-x}\right),~~&~0<x<\frac{1}{2}\\ (-1)^{p-1}\mathrm{Li}_{1-p}\left(1-\frac{1}{x}\right),~~&~ \frac{1}{2}<x<1\end{cases}\end{split} \label{eq:cumulant_series.1}
    \end{align}
    Here, $\mathrm{Li}_{s}(x)$ is the polylogarithm function, defined as $\mathrm{Li}_{s}(x)=\sum_{k=1}^{\infty}k^{-s}x^{k}$.
    From Eq.~\eqref{eq:asymptotic_eigenvalue}, in the limit of a large number of
    particles $N$, the overlap-matrix eigenvalues can be approximated by the
    complementary error functions. The limit allows us to convert sums to integrals
    and write the cumulants in the form,
    \begin{equation}
        \langle \hat{N}_{r}^{p}\rangle_{c}\approx (2\pi r)\,\mathcal{J}^{p}_{N}(\mu
        ;r),
    \end{equation}
    where,
    \begin{align}
        \mathcal{J}^{p}_{N} & (\mu;r)=\frac{1}{\pi}\int^{\frac{(\mu+1)N}{2r}-\frac{r}{2}}_{\frac{\mu N}{2r}-\frac{r}{2}}\diff{z}\label{eq:cumulant_integral_0}                                                                                \\
                            & \times ~\left[n_{p}\left(\tfrac{1}{2}{\rm erfc}\left(z\sqrt{2}
    \!\left(1+\tfrac{2z}{r}\right)^{\!-1/2}\right)\right)\right],\nonumber
    \end{align}
    \noindent
    where $n_{p}(x)$ is defined in Eq.~\eqref{eq:cumulant_series.1}. As in the case of the R\'enyi entropy in Sec.~\ref{section:disc_entropy}, we find that the cumulants possess different scaling forms in the three
    regimes for large $N$. We find,
    \begin{align}
         & \frac{\left\langle \hat{N}_{\sqrt{N}\xi}^{p}\right\rangle_{c}}{\sqrt{N}}\approx 2\pi \xi \label{eq:N^p-scaling}                                                                                                                                                                              \\
         & \times~ \begin{cases}j_{p}\left(\sqrt{N}(\xi-\sqrt{\mu})\right)&|\xi-\sqrt{\mu}|\sim\frac{1}{\sqrt{N}},\\[4pt] ~~~~~\kappa_{p};~&\sqrt{\mu}\ll \xi \ll \sqrt{1+\mu},\\[4pt] j_{p}\left(-\sqrt{N}(\xi-\sqrt{1+\mu})\right);~&|\xi-\sqrt{1+\mu}| \sim \frac{1}{\sqrt{N}},\end{cases} \nonumber
    \end{align}
    where
    \begin{equation}
        j_{p}(y)=[\sqrt{2}\pi]^{-1}\int_{-\infty}^{y\sqrt{2}}\diff{z}\, n_{p}\left(\tfrac{1}{2}{\rm erfc}(z)\right), \label{eq:cumulant_scaling_0}
    \end{equation}
    Since, $n_p(x)$ given in Eq.~\eqref{eq:cumulant_series.1} is a bounded function for $0 \le x \le 1$, the above integral $j_{p}(y)$ converges for $y \to \infty$ to a finite value,
    \begin{align}
        \kappa_{p}= j_{p}(\infty), \label{def:kappa_p}
    \end{align}
    for $p\ge 2$.
    \begin{figure}[ht!]
        \centering
        \includegraphics[width=\columnwidth]{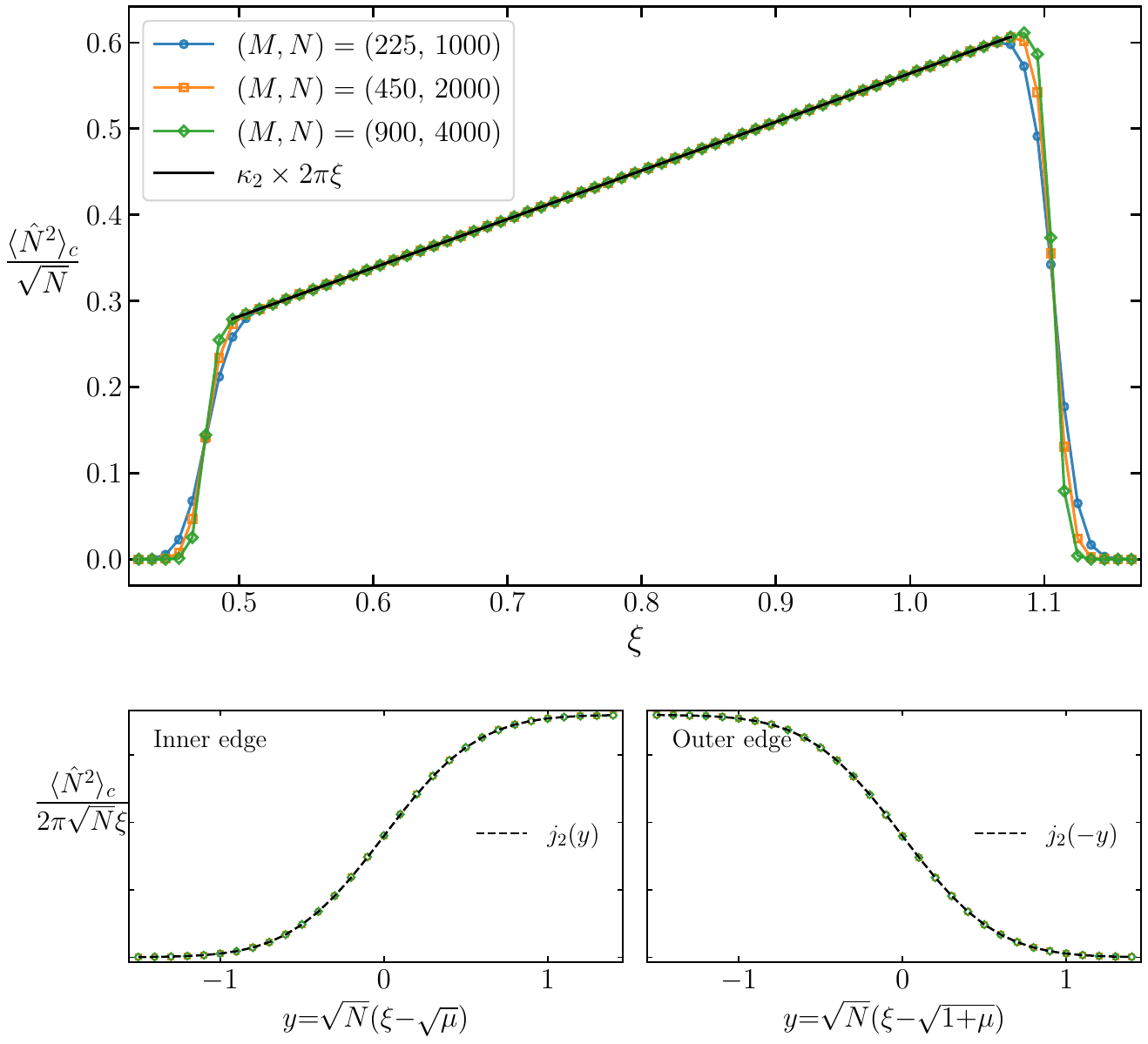}
        \caption{The upper panel shows the variance $\la \hat{N}^{2}\ra_{c}$ as a function of the rescaled radial coordinate $\xi$, for several values of $(M,N)$ at fixed $\mu=0.225$. Linear growth in the \textit{bulk} region is confirmed, with area-law coefficient $\kappa_{2}=j_2(\infty)=\frac{1}{2\pi^{3/2}}$. The lower panel compares the numerically computed ratio $\la \hat{N}^{2}\ra_{c}/(2\pi\sqrt{N}\xi)$ (marked points) with the scaling function $j_{2}$ of Eq.~\eqref{eq:N^p-scaling} (dashed lines) near the inner and outer edges.}
        \label{fig:variance_zeta}
    \end{figure}
    
    Numerical analysis of Eq.~\eqref{eq:cumulant_scaling_0} shows that the odd cumulants vanish in the \textit{bulk} region, consistent with the inference from the \textit{bulk} CCGF in Eq.~\eqref{eq:ccgf_bulk}. For $p=2$ one finds $\kappa_{2}=\tfrac{1}{2\pi^{3/2}}$, in agreement with the corresponding ground-state result reported in Ref.~\cite{PhysRevA.99.021602}. In figure \ref{fig:variance_zeta}, we have numerically checked the scaling relation \eqref{eq:N^p-scaling}
    for the variance of the disc.

    \begin{figure}[ht!]
        \centering
        \includegraphics[width=\columnwidth]{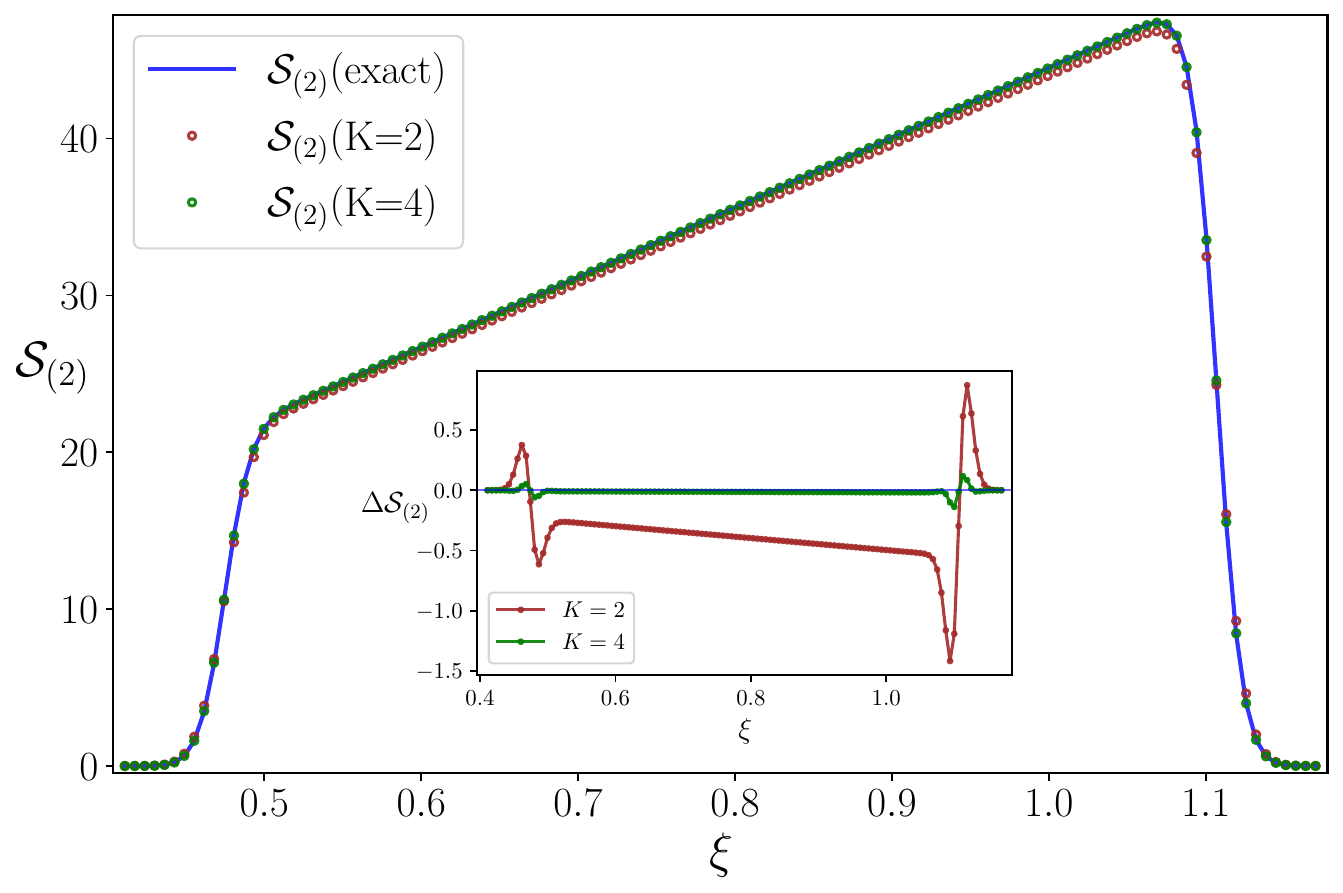}
        \caption{A comparison of the second R\'enyi entropy $\s_{(2)}$, as a function of the rescaled radius $\xi$ of the disc region (blue line) with cumulant expansions (Eq.~\eqref{eq:renyi_cumulant_relation}) truncated at orders $K=2$ (brown circles) and $K=4$ (green circles). The cumulant expansion shows satisfactory convergence to the exact result even at the lowest truncation order $K=2$. The inset shows the difference $\s_{(2)}(K) - \s_{(2)}({\mathrm{exact}})$, confirming improvement with increasing $K$.}
        \label{fig:renyi_cumulant_sum}
    \end{figure}

    We now verify the relation in Eq.~\eqref{eq:renyi_cumulant_relation}, which connects the entanglement entropy (EE) to the full counting statistics (FCS) in the disc region $\mathcal{D}_{r}$. To this end, we compare two quantities: (i) $\s_{2}$, computed numerically by inserting the exact overlap-matrix eigenvalues from Eq.~\eqref{eq:disc_eigenvalues} into the R\'enyi entropy expression in Eq.~\eqref{eq:renyi_entropy_overlap_matrix}, and (ii) $\s_{2}^{(K)}$, the sum on the right-hand side of Eq.~\eqref{eq:renyi_cumulant_relation} truncated at $p=K$. The coefficients $\beta_{p}(q)$ appearing in Eq.~\eqref{eq:renyi_cumulant_relation} are given by~\cite{PhysRevB.83.161408,song2012bipartite}
    \begin{equation}
    \beta_{p}(q)=\begin{cases}
        \dfrac{2}{q-1}\dfrac{1}{p!}\left(\dfrac{2\pi i}{q}\right)^{p}
        \zeta\!\left(-p,\dfrac{q+1}{2}\right), & \text{even }p, \\[0.6em]
        0,                                      & \text{odd }p,
    \end{cases}
    \label{eq:beta_infinite}
    \end{equation}
    where $\zeta(s,a)$ denotes the Hurwitz zeta function. As shown in Figure~\ref{fig:renyi_cumulant_sum}, the truncated cumulant expansion converges to the exact EE as $K$ increases, providing a direct numerical verification of Eq.~\eqref{eq:renyi_cumulant_relation}.

\begin{figure}
        \centering
        \includegraphics[width=\columnwidth]{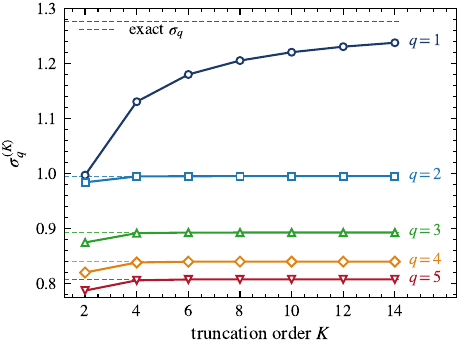}
        \caption{
        Convergence of the cumulant series for the R\'enyi area-law coefficient. Symbols show the truncated sum $\sigma_q^{(K)}=\sum_{p=2,4,\dots}^{K}\beta_p(q)\,\kappa_p$ [Eq.~\eqref{eq:entropy_cumulant_relation_1}] as a function of the truncation order $K$, for $q=1,\dots,5$; dashed lines are the exact coefficients $\sigma_q$ obtained by direct integration of LHS of Eq.~\eqref{eq:integral_identity_1}. For $q\ge2$ the series converges rapidly, reaching a relative accuracy of $10^{-10}$ ($q=2$) to $10^{-8}$ ($q=5$) at $K=14$. The von Neumann case $q=1$ converges only marginally---and is still $3\%$ below $\sigma_1$ at $K=14$.}
    \label{fig:truncated_sum}
    \end{figure}

    This relation also leads to an interesting mathematical identity. In the bulk region, both the entropy and the number cumulants grow linearly with $r$ at leading order, $\mathcal{S}_{q}\sim \sigma_{q}\times(2\pi r)$ and $\la \hat{N}^{p}_{r} \ra_{c}\sim \kappa_{p}\times(2\pi r)$. Substituting these scalings into Eq.~\eqref{eq:renyi_cumulant_relation} shows that the bulk area-law coefficients $\sigma_{q}$ and $\kappa_{p}$, defined in Eqs.~\eqref{eq:sigma_q} and~\eqref{def:kappa_p}, satisfy
    \begin{equation}
    \sigma_{q}=\sum_{p=1}^{\infty}\beta_{p}(q)\,\kappa_{p}.
    \label{eq:entropy_cumulant_relation_1}
    \end{equation}
    Written out explicitly,  the integral identity is 
    \begin{align}
     & (1-q)^{-1}\int_{-\infty}^{\infty}\diff{z}\,\ln\!\left[
       \left(\tfrac{1}{2}\mathrm{erfc}(-z)\right)^{q}+
       \left(\tfrac{1}{2}\mathrm{erfc}(z)\right)^{q}\right]\nonumber \\
     & =\sum_{p=1}^{\infty}\beta_{2p}(q)\int_{-\infty}^{\infty}
       \diff{z}\,n_{2p}(\tfrac{1}{2}\mathrm{erfc}(z)),\label{eq:integral_identity_1}
    \end{align}
    where $n_{2p}(x)$ is given in Eq.~\eqref{eq:cumulant_series.1}. In Table.~\ref{tab:entropy_cumulant_verification} and  Fig.~\ref{fig:truncated_sum}, we show a numerical verification of the above identify. The truncated sums of the area law coefficients of the even-cumulant is seen to converge fast to the R\'enyi area-law coefficients, except for the $q=1$ case (corresponding to the von-Neumann entropy)  where the convergence is very slow.
    \begin{table}[ht!]
    \centering
    \setlength{\tabcolsep}{5pt}
    \renewcommand{\arraystretch}{1.45}
    \begin{tabular}{| c | c | c | c | c | c |}
    \hline
    & & \multicolumn{4}{c|}{$\sum_{p=2,4,\dots}^{K}\beta_{p}(q)\,\kappa_{p}$} \\
    \cline{3-6}
    $q$ & $\sigma_{q}$ & $K=2$ & $K=4$ & $K=6$ & $K=8$ \\
    \hline
    $1$ & $1.27731$ & $0.997356$ & $1.13090$ & $1.18041$ & $1.20569$ \\
    $2$ & $0.995443$ & $0.984351$ & $0.995053$ & $0.995426$ & $0.995442$ \\
    $3$ & $0.893027$ & $0.874978$ & $0.891891$ & $0.892938$ & $0.893019$ \\
    $4$ & $0.840149$ & $0.820292$ & $0.838576$ & $0.839993$ & $0.840132$ \\
    $5$ & $0.807899$ & $0.787480$ & $0.806111$ & $0.807700$ & $0.807875$ \\
    \hline\hline
    \end{tabular}
    \caption{Verification of the identity in Eq.~\eqref{eq:integral_identity_1}, for $q=1, 2,\dots,5$. The right hand side is computed by truncating the series at $p=K$. For the case of $q=1$, the convergence of the sum to $\sigma_1$ is much slower. For a pictorial demonstration of the convergence, see Fig.~\ref{fig:truncated_sum}.}
    \label{tab:entropy_cumulant_verification}
    \end{table}
    \subsection{Distribution of $\hat N_r$}\label{sec:ldf}
    \noindent
    In this section we discuss the behaviour of the distribution of $\hat N_r$ for large $M,N$ (with $\mu=M/N$ fixed) in the bulk regime: $\sqrt{\mu}<\xi <\sqrt{1+\mu}$.
    
    From Eqs.~\eqref{<N>-scaling} and \eqref{eq:N^p-scaling}, we recall that $\langle \hat N_r \rangle \approx N (\xi^2-\mu)$ and $\langle \hat N_r^p \rangle_c \approx 2 \pi \xi~\sqrt{N} \kappa_p$ for $p\ge 2$, in the bulk regime $\sqrt{\mu} < \xi < \sqrt{1+\mu}$. This indicates that the distribution of the scaled random variable $\frac{\hat N_r - \langle \hat N_r \rangle}{\sqrt{\langle \hat N_r^2 \rangle_c}}$ converges to a  Gaussian distribution of mean zero and unit variance. This implies that the typical fluctuations of $\varsigma =\frac{\hat N_r}{N}$ around its mean $\xi^2-\mu$ are zero-mean Gaussian random variables of order $\sim N^{-3/4}$. We now investigate the fluctuations of order much larger than this typical scale.
    
    From the definition of the CCGF 
        \begin{align}
            \chi(\lambda;r)&= \ln\left\langle e^{-\lambda (\hat{N}_r -\langle \hat N_r \rangle)}\right \rangle \notag \\
            &= \ln\left[\sum_{K=0}^{N}e^{-\lambda (K-\langle K \rangle)}P_{r}(K;M,N)\right], \label{def:Z-with-P}
        \end{align}
    in Eq.~\eqref{ccgf_logarithm}, one can obtain $P_r(K;M,N)$, the probability  of finding $K$ particles inside the disc of radius $r=\sqrt{N}\xi$, in the excited state $\ket{\psi_{(M,M+N)}}$. This can be done by performing an inverse Laplace transformation,
    \begin{align}
    P_r(K;M,N) = \int \frac{d\lambda}{2\pi i}~ e^{\lambda(K-\la K \ra) + \chi(\lambda;r)}.
    \end{align}
    Using the definitions from Eqs.~\eqref{eq:cumgf} and ~\eqref{def:<N_r>} in Eq.~\eqref{ccgf_logarithm}, the CCGF $\chi(\lambda;r)$ can be written explicitly as a sum over the overlap-matrix eigenvalues $a_{s}(r)$,
    \begin{equation}
        \chi(\lambda;r)=\sum_{s=1}^{N}\left\{\ln\left[1+(\mathrm{e}^{-\lambda}-1)\,a_{s}(r)\right]+\lambda\, a_{s}(r)\right\}.
        \label{eq:ccgf_sum}
    \end{equation}
    In the large-$N$ limit we replace these eigenvalues by their asymptotic form in Eq.~\eqref{eq:asymptotic_eigenvalue} and change variables from $s$ to $z=\frac{M+s-r^{2}}{\sqrt{2}\,r}$, as in Sec.~\ref{subsec:disc_entanglement}. Since $r\sim O(\sqrt{N})$, the increment $\diff{z}=\frac{1}{\sqrt{2}\,r}$ is small and the sum in Eq.~\eqref{eq:ccgf_sum} converges to the integral
    \begin{equation}
    \label{eq:ccgf_largen}
    \begin{split}
            \chi(\lambda;r) =& \sqrt{2}r \int_{-\frac{r}{\sqrt{2}}+\frac{\mu N}{\sqrt{2}r}}
            ^{\frac{(\mu+1)N}{\sqrt{2}r}-\frac{r}{\sqrt{2}}}\diff{z}\\
            &\left[ \ln \left
            (1+\frac{e^{-\lambda}-1}{2}\mathrm{erfc}\!\left(z\left(1+\frac{\sqrt{2}z}{r}
            \right)^{-1/2}\right)\right)\right.\\
            &~~\left.+\frac{\lambda}{2}\mathrm{erfc}\!\left(z\left(1+\frac{\sqrt{2}z}{r}\right)^{-1/2}\right)
            \right].
    \end{split}
    \end{equation}
    In the \textit{bulk} regime, $\sqrt{\mu}< \xi < \sqrt{1+\mu}$ with $r=\sqrt{N}\xi$, performing a similar large-$N$ approximation as applied to $\mathcal{I}_{q}$ in Appendix~\ref{sec:bulk_renyi} reduces Eq.~\eqref{eq:ccgf_largen} to
    \begin{align}
        \begin{split}&\chi(\lambda;r)=\sqrt{2}r\\&~\times~ \int_{-\infty}^{\infty}\diff{z}\left[ \ln \left(1+\frac{e^{-\lambda}-1}{2}\mathrm{erfc}\!\left(z\right)\right) +\lambda\frac{\mathrm{erfc}\!\left(z\right)}{2}\right]\end{split}
        \label{eq:ccgf_largen_1}
    \end{align}
    We split the integration range into two parts, $z\in[0,\infty)$ and $z\in(-\infty,0]$ and make the transformation $z\to-z$ in the second part. Then, using the identity $\mathrm{erfc}(z)+\mathrm{erfc}(-z)=2$, the CCGF takes the form,
    \begin{align}
        \begin{split}\chi(\lambda;\sqrt{N} \xi) \approx&\sqrt{2N}\xi \Tilde{\chi}(\lambda), ~ \text{where,}~ \\ \Tilde{\chi}(\lambda)=&\int_{0}^{\infty}\diff{z}\ln\left[ 1+\sinh^{2}\left(\frac{\lambda}{2}\right) \mathrm{erfc}(z)\mathrm{erfc}(-z) \right].\end{split} \label{eq:ccgf_bulk}
    \end{align}
    The resulting scaling function $\Tilde{\chi}(\lambda)$ is an even function of $\lambda$, implying that all odd cumulants in the \textit{bulk} vanish in this limit. This form is identical to the CCGF obtained for the corresponding problem of number fluctuations in the Ginibre ensemble~\cite{PhysRevE.100.012137}. Its asymptotic behavior, obtained there, reads
    \begin{equation}
        \Tilde{\chi}(\lambda)\approx
        \begin{cases}
            \dfrac{\lambda^{2}}{2\sqrt{2\pi}}, & \lambda\to 0,          \\[6pt]
            \dfrac{2}{3}|\lambda|^{3/2},       & \lambda\to \pm \infty.
        \end{cases}
        \label{eq:tchi(lambda)}
    \end{equation}
    Using the scaling behaviours of $\chi(\lambda;\xi)$ and $\langle \hat{N}_{r}\rangle$ from Eqs.~\eqref{eq:ccgf_bulk} and \eqref{<N>-scaling} in Eq.~\eqref{ccgf_logarithm}, we see that the MGF $\mathcal{Z}(\lambda;r)$ has the following asymptotic form
    \begin{align}
        \mathcal{Z}(\lambda;\sqrt{N}\xi) \approx \exp \left(N\lambda(\mu-\xi^{2}) +\sqrt{2N}\xi\tilde{\chi}(\lambda) \right),
        \label{eq:mcalZ}
    \end{align}
    for large $N$ in the bulk region $\sqrt{\mu}< \xi <\sqrt{(1+\mu)}$. Performing an inverse Laplace transformation of $\mathcal{Z}(\lambda;r)$ in Eq.~\eqref{eq:mcalZ}, we get the distribution $P_{r}(K;M,N)$. For large $N$ in the bulk region we find that the probability $P_r(K;M,N)$ has the following scaling property:
    \begin{equation}
        P_{r}(K;\mu,N) =\frac{1}{N}\mathcal{P}_{\frac{r}{\sqrt{N}}}\left(\frac{K}{N}
        ;\mu,N\right),
    \end{equation}
    where $\varsigma=K/N$ and,
    \begin{equation}
        \mathcal{P}_{\xi}(\varsigma; \mu,N) \approx \int_{\mathcal{C}}\frac{\diff{\lambda}}{2\pi i}\, \e^{N \lambda (\varsigma -\xi^{2}+\mu)+\sqrt{2N}\xi\Tilde{\chi}(\lambda)}.\label{eq:inverse_laplace_pdf}
    \end{equation}
    Note that, setting  $\frac{\varsigma-\xi^{2}+\mu}{\xi\sqrt{2}}=\frac{\varphi}{\sqrt{N}}$ with $\varphi \sim O(1)$, makes the two terms in the exponent of the integrand of Eq.~\eqref{eq:inverse_laplace_pdf}, same order {\it i.e.,} $O(\sqrt{N})$.  Consequently, for large $N$, one can perform the integral by a saddle point approximation which yields the following scaling form,
    \begin{equation}
        \mathcal{P}_{\xi}(\varsigma; \mu,N)\approx e^{-\sqrt{2N}\xi \, \Psi\left[\sqrt{\frac{N}{2\xi^{2}}}(\varsigma-\xi^2+\mu)\right]},\label{eq:pdf_scaling form}
    \end{equation}
    where the rate function $\Psi(\varphi)$ is given by,
    \begin{equation}
        \Psi(\varphi)=-\min_{\lambda}~\lambda\varphi+\tilde{\chi}(\lambda).\label{eq:pdf_rate_function}
    \end{equation}
    Using the asymptotic behaviour of $\tilde{\chi}(\lambda)$ in Eq.~\eqref{eq:tchi(lambda)}, one can show that the  rate function $\Psi(\varphi)$ asymptotically goes as ~\cite{PhysRevE.100.012137}, 
    \begin{align}
    \Psi(\varphi) \sim 
        \begin{cases}
        \sqrt{\frac{\pi}{2}} \varphi^2,~&~\varphi \to 0, \\
        \frac{1}{3} |\varphi|^3,~&~\varphi \to \infty.
        \end{cases}
    \label{eq:Psi(Phi)-asym}
    \end{align}
    The scaling form of Eq.~\eqref{eq:pdf_scaling form} describes $O(N^{-1/2})$ fluctuations of $\varsigma$ around its mean which are larger than the typical fluctuations of order $O(N^{-3/4})$.  This scaling form is valid within the \textit{bulk} regime, $\sqrt{\mu}< \xi < \sqrt{\mu+1}$  for $|\varsigma -(\xi^2-\mu)| \sim O(N^{-1/2})$. In this regime, the rate function~\eqref{eq:pdf_rate_function} coincides with that of the state $\ket{\psi_{(0,N)}}$ of $N$ rotating fermions in a two-dimensional harmonic trap \cite{PhysRevE.100.012137}, up to a shift of the variable $\varsigma$ by $\mu$.
    
    The distribution of $\hat N_{r}$ in the $M=0$ state has  previously been analysed in~\cite{allez2014index, shirai2006large,PhysRevE.100.012137}. These studies reveal that the distribution of number fluctuations shows three regimes --(i) typical [$|\varsigma-\xi^{2}| \sim O(N^{-3/4})$], (ii) intermediate [$|\varsigma-\xi^{2}| \sim O(N^{-1/2})$] and (iii) large deviation [$|\varsigma-\xi^{2}| \sim O(N^0)$]. It was also shown in \cite{PhysRevE.100.012137} that the rate function $\Psi(\varphi)$ in Eq.~\eqref{eq:pdf_scaling form} describes the number fluctuations only in the intermediate regime and the two asymptotic limits in Eq.~\eqref{eq:Psi(Phi)-asym} smoothly connect the typical and large deviation regimes. However, the scaling analysis in Eqs.~\eqref{eq:pdf_rate_function} - \eqref{eq:Psi(Phi)-asym} is valid only for the intermediate regime and  does not work in the large deviation regime. This is because the fluctuations in this regime depart from the mean by an extensive amount ($O(1))$. One then needs to follow a different procedure,  as described in \cite{PhysRevE.100.012137}. Following the same procedure for the 
    excited state, we find that the distribution of the fluctuations in the atypical regime $|\varsigma-(\xi^2-\mu)|\sim O(1)$, takes the large deviation form, \cite{allez2014index,PhysRevE.100.012137},
    \begin{equation}
    \label{eq:LDexc}
      {\cal P}_\xi(\varsigma;\mu,N)\approx e^{-N^2\Psi_{\rm ld}(\varsigma)},
    \end{equation}
    with the rate function,
    \begin{align}
      &\Psi_{\rm ld}(\varsigma)=\frac{\operatorname{sgn}(\xi^2-\varsigma-\mu)}{4},\label{eq:Psiexc}\\
      &~~\times~\left[\xi^4-4\xi^2(\varsigma+\mu)+3(\varsigma+\mu)^2+2(\varsigma+\mu)^2\ln\tfrac{\xi^2}{\varsigma+\mu}\right], \notag
    \end{align}
    valid for $0\le\varsigma\le1$. Note that, for $|\varsigma-(\xi^2-\mu)| \to 0$, the function $\Psi_{\rm ld}(\varsigma)$ behaves asymptotically as 
    \begin{align}
    \Psi_{\rm ld}(\varsigma) \approx
    \frac{|\varsigma-\xi^2+\mu|^3}{6\xi^2},
    \end{align}
    which clearly smoothly connects to the large argument asymptotic behaviour of the intermediate regime in Eq.~\eqref{eq:Psi(Phi)-asym}.
    
    Combining the behaviour of the distribution of the number fluctuation in different fluctuation scales, we, in summary, have
        \begin{align}
    \label{eq:ground_pdf}
    \mathcal{P}_{\xi}&(\varsigma;\mu,N) \approx \notag \\
     &   \begin{cases}
            e^{-\frac{N^{\frac{3}{2}}\sqrt{\pi}}{2\xi}\left(\varsigma_\mu-\xi^{2}\right)^{2}},
            & |\varsigma_\mu-\xi^{2}|\sim \frac{1}{N^{\frac{3}{4}}},\\[6pt]
            e^{-\sqrt{2N}\xi\Psi\!\left[\sqrt{\frac{N}{2\xi^{2}}}(\varsigma_\mu-\xi^{2})\right]},
            & |\varsigma_\mu-\xi^{2}|\sim \frac{1}{N^{\frac{1}{2}}},\\[6pt]
            e^{-N^{2}\Psi_{\rm ld}(\varsigma_\mu)},
            & |\varsigma_\mu-\xi^{2}|\sim N^{0},
        \end{cases}
    \end{align}
    where $\varsigma_\mu=\varsigma+\mu$, with $\Psi(\varphi),~\Psi_{\rm ld}(\varsigma)$ given by Eqs.~\eqref{eq:Psi(Phi)-asym} and \eqref{eq:Psiexc}.
    
    \section{Entanglement and FCS of annular region}
    \label{sec:annulus}

    We next consider the Entanglement Entropy (EE) and full counting statistics
    (FCS) for an annular region $\mathcal{A}_{r_{1},r_{2}}$, bounded by two concentric
    circles of radii $r_{1}$ and $r_{2}$. Assuming the condition, $M \ll r_{1}^{2}
    \ll r_{2}^{2}\ll M+N$ ensures that the annulus lies within the \textit{bulk}
    of the excited state (see Fig.~\ref{fig:particle_density_profile}).
    Eq.~\eqref{eq:annulus_eigenvalues_0} shows that the annulus eigenvalues $\Tilde
    {a}_{s}(r_{1},r_{2})$ are given by the difference between the eigenvalues for
    the two bounding discs:
    \begin{equation}
        \Tilde{a}_{s}(r_{1},r_{2})=a_{s}(r_{2})-a_{s}(r_{1}), \label{eq:annulus_eigenvalues}
    \end{equation}
    Substituting these annulus eigenvalues into the moment generating function (MGF) for the particle number, $\hat N_{r_1,r_2}$, in the annular region ${\cal A}_{r_1,r_2}$, from Eq.~\eqref{eq:cumgf}, we have,
    \begin{equation}
        \Z(\lambda; r_{1},r_{2}) =\prod_{s=1}^{N}\!\left[1-\Tilde{a}_{s}(r_{1},r_{2}
        )+\e^{-\lambda}\Tilde{a}_{s}(r_{1},r_{2})\right],
    \end{equation}
    where $\mathcal{Z}(\lambda; r_{1}, r_{2})\equiv {\cal Z}(\lambda; {\cal A}_{r_1, r_2})$.
    The corresponding centred cumulant generating function (CCGF), is then,
    \begin{equation}
    \begin{split}
        \chi(\lambda; r_{1},r_{2}) =&\ln \la e^{-\lambda (\hat N_{r_1,r_2}-\la \hat N_{r_1,r_2} \ra)} \ra\\
        =&\ln \mathcal{Z}(\lambda; r_1,r_2)+\lambda \la \hat{N}_{r_{1},r_{2}}\ra ,
    \end{split}\label{ccgf_logarithm_annulus}
    \end{equation}
    where $\la \hat{N}_{r_1,r_2}\ra = \sum_{s=1}^{N}\tilde{a}_{s}(r_{1},r_{2})$ is the mean number of particles in the annulus ${\cal A}_{r_1,r_2}$. Therefore,
    \begin{equation}
    \begin{split}
    \chi(\lambda;r_{1},r_{2}) =&\sum_{s=1}^{N}~\ln \left[1-a_{s}(r_{2})+a_{s}(r_{1})+\e^{-\lambda}a_{s}(r_{2})\right.\\
    &\left.-\e^{-\lambda}a_{s}(r_{1})\right]+\lambda \sum_{s=1}^{N}\{a_{s}(r_{2})-a_{s}(r_{1})\}.
    \end{split}
    \label{eq:ccgf_log_annulus1}
    \end{equation}
    
    \begin{figure}[ht!]
        \centering
        \includegraphics[width=\columnwidth]{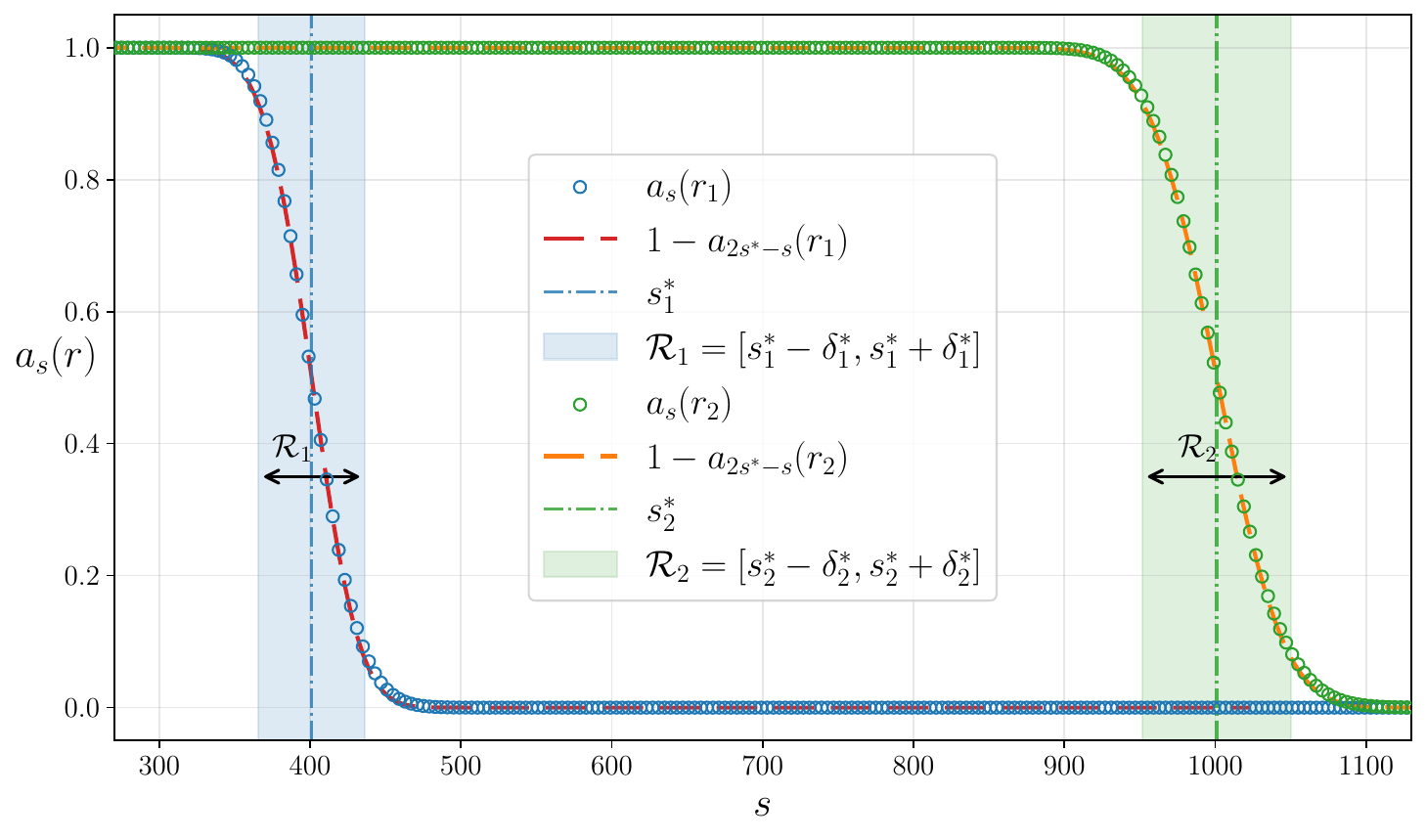}
        \caption{Numerical verification of the reflection symmetry relation $a_{s}(r) \approx 1 - a_{2s^* - s}(r)$ for the eigenvalues $a_{s}(r)$. The circles represent the eigenvalues
        $a_{s}(r)$ as a function of the index $s$ for $M=225$,
        evaluated at radii $r_{1}=25$ (blue) and $r_{2}=35$ (green). The red and orange dashed lines represent $1 - a_{2s^* - s}(r)$ at $r_1$ and $r_2$ respectively. The reflection
        symmetry relation is confirmed by the perfect overlap of the circles with the dashed lines. Vertical dot-dashed lines
        indicate the crossover points $s^{*}= r^{2}- M + 1$, where the eigenvalues
        transition from 1 to 0. The shaded areas highlight the transition regions
        $\mathcal{R}= [s^{*}- \delta^{*},s^{*}+\delta^{*}]$, illustrating the
        characteristic crossover width $\delta^{*}=\sqrt{2(M+s^{*})}$.}
        \label{fig:eigenvalue_k}
    \end{figure}

    The disc eigenvalues $a_{s}(r)$ for large $N,M$ have the asymptotic form given
    by $a_{s}(r)\approx \frac{1}{2}\mathrm{erfc}(y),\textrm{where}\,\,y=\frac{M+s-1-r^{2}}{\sqrt{2(M+s-1)}}$
    as given in Eq.~\eqref{eq:asymptotic_eigenvalue}. As a function of increasing $s$, the eigenvalues $a_s(r)$ transition from 1 to 0, as one passes through $s^{*}=r^{2}-M+1$. This occurs over a crossover region of width $\delta^{*}\approx\sqrt{2(M+s^{*})}$. Corresponding to the two radii $r_{1}$ and $r_{2}$,
    we get the crossover points $s_{1}^{*}\equiv r_{1}^{2}-M+1$ and $s_{2}^{*}\equiv r_{2}^{2}-M+1$ respectively. We consider
    the case where the boundaries of the discs are well separated:
    $(r_{2}^{2}-r_{1}^{2})\gg \mathcal{O}(N^{1/2})$. For this choice, the sums
    in Eq.~\eqref{eq:ccgf_log_annulus1} get significant constribution from the
    following two regions -- $\mathcal{R}_{1}=[s_{1}^{*}-\delta_{1}^{*},s_{1}^{*}
    +\delta_{1}^{*}]$ and $\mathcal{R}_{2}=[s_{2}^{*}-\delta_{2}^{*},s_{2}^{*}+\delta
    _{2}^{*}]$. Note that in the first range $a_{s}(r_{2})\approx 1$, whereas in
    the second range $a_{s}(r_{1})\approx 0$. Using these facts, we get,
    \begin{align}
        \chi(\lambda;r_{1},r_{2})\nonumber  \\
        \approx \sum_{s \in \mathcal{R}_1} & \ln\!\left[a_{s}(r_{1})+e^{-\lambda}(1-a_{s}(r_{1}))\right]+\lambda (1-a_{s}(r_{1}))\nonumber              \\
        +\sum_{s \in \mathcal{R}_2}        & \ln\!\left[1-a_{s}(r_{2})+e^{-\lambda}a_{s}(r_{2})\right]+\lambda a_{s}(r_{2}). \label{eq:log_ccgf_approx}
    \end{align}
    The identity $\mathrm{erfc}(y)+\mathrm{erfc}(-y)=2$ implies the reflection symmetry
    $a_{2s^*_1-s}(r_{1})\approx1-a_{s}(r_{1})$ (see Fig.~\ref{fig:eigenvalue_k}).
    Therefore, we make the transformation $s'=2s_{*}-s$. Noting that the range
    of summation remains unchanged, we obtain,
    \begin{align}
        \chi(\lambda;r_{1},r_{2})\nonumber  \\
        \approx \sum_{s \in \mathcal{R}_1} & \ln\!\left[1-a_{s}(r_{1})+e^{-\lambda}a_{s}(r_{1})\right]+\lambda
        a_{s}(r_{1})\nonumber                           \\
        +\sum_{s \in \mathcal{R}_2}        & \ln\!\left[1-a_{s}(r_{2})+e^{-\lambda}a_{s}(r_{2})\right]+\lambda a_{s}(r_{2})\label{eq:ccgf_annulus_disc_relation}
    \end{align}
    We can extend the range of summations in Eq.~\eqref{eq:ccgf_annulus_disc_relation}
    to the full range $s=\{1,\dots,N\}$ because $a_{s}$ is either 0 or 1 in the
    added regions. Hence, comparing with Eq.~\eqref{ccgf_logarithm}, we have,
    \begin{equation}
        \chi(\lambda;r_{1},r_{2})\approx \chi(\lambda;r_{1})+\chi(\lambda;r_{2}).
    \end{equation}
    This result demonstrates the additivity of the CCGF for the annulus. Consequently,
    the cumulants of the particle number distribution satisfy
    \begin{equation}
        \la \hat{N}^{p}_{r_{1},r_{2}} \ra_{c}\approx \la \hat{N}^{p}
        _{r_{1}}\ra_{c}+ \la \hat{N}^{p}_{r_{2}}\ra_{c}~~\text{for}~~p\ge 2.
    \end{equation}
    Finally, given the correspondence between FCS and entanglement entropy \bluew{(Eq.~\eqref{eq:renyi_cumulant_relation})}, the
    Rényi entropies for the annulus also decompose additively:
    \begin{equation}
        \s_{q}(\mathcal{A}_{r_{1},r_{2}})\approx \s_{q}(\mathcal{D}_{r_{1}})+\s_{q}
        (\mathcal{D}_{r_{2}}).\label{eq:annulus_disc_ee}
    \end{equation}
    Alternatively, we can arrive at Eq.~\eqref{eq:annulus_disc_ee} starting from
    the definition of $\s_{q}$ from Eq.~\eqref{eq:renyi_entropy_overlap_matrix}
    and using a similar analysis.
    \begin{figure}[htb!]
        \includegraphics[width=\columnwidth]{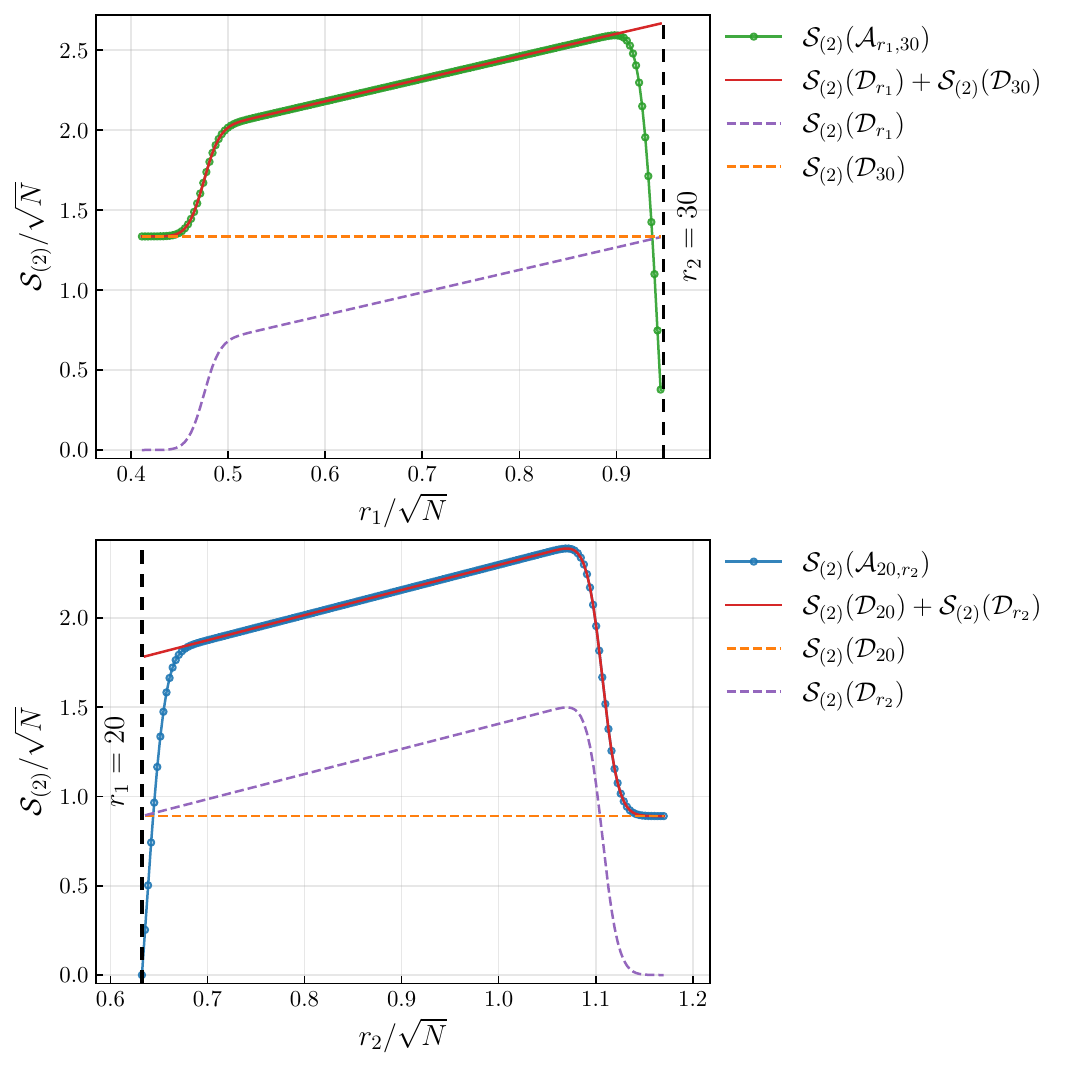}
        \caption{Verification of the additivity relation Eq.~\eqref{eq:annulus_disc_ee} for R\'enyi entropy $\s_{(2)}$, for fixed $r_1$ (top) and $r_2$ (bottom), with $M=225, N=1000$. The vertical dashed lines represent the locations of $r_1$ and $r_2$ respectively. The circles are obtained by using the eigenvalues from Eq.~\eqref{eq:annulus_eigenvalues} in Eq.~\eqref{eq:renyi_entropy_overlap_matrix}. The dashed lines are $\s_{(2)}$ of the bounding discs at radius $r_1$ and $r_2$ obtained using the eigenvalues from Eq.~\eqref{eq:disc_eigenvalues} in Eq.~\eqref{eq:renyi_entropy_overlap_matrix}. The solid red lines represent $\s_{q}(\mathcal{D}_{r_{1}})+\s_{q}
        (\mathcal{D}_{r_{2}})$. In both panels, we observe excellent verification of the additivity relation except when $|r_2-r_1| \le O(1)$.}
        \label{fig:annulus_entropy_comparison}
    \end{figure}
    \section{Summary and Outlook}
    \label{section:conclusion} 
    It is usually expected that the many-body ground state of a gapped non-interacting fermionic system will exhibit area law entanglement, while high temperature states show volume law entanglement~\cite{RevModPhys.80.517}. In this work, we provided a simple example for the case of non-interacting fermions where we construct pure states at  arbitrarily high energies (high temperatures) for which we show that the area law entanglement scaling continues to hold.
    
    We considered the set up of $N$ non-interacting fermions in a rotating two-dimensional harmonic trap. The exact single particle spectrum is known for this system and one can construct many-body slater determinant states  by filling any $N$ single particle levels. The state under consideration was chosen by filling $N$ single-particle levels, given by Eq.~\eqref{eq:energy_spectrum}, in the $l=0$ sector with $k=M,\dots,M+N-1$, where $M\ge 0$. This corresponds to filling single-particle states with different angular momentum. For this state, the particle density falls rapidly to zero outside the domain $\sqrt{M}<r<\sqrt{M+N}$, for large $M$ and $N$. We then derived exact analytical expressions for the $q$th R\'enyi entropies and particle number cumulants of a specified domain (a disc and an annulus).
    
    For the disc geometry, we observed that the entanglement entropy and cumulants scale linearly with the circumference of the disc, as long as the boundary of the disc is in the bulk region, where the particle density is constant. This establishes, for our model, the area law scaling for the entanglement entropy for a set of high energy states. The coefficients for the linear dependence were computed explicitly. The resulting centered cumulant-generating function and its associated probability distribution are found to be identical, up to a variable shift, to the corresponding problem for the $M=0$ state, studied previously in~\cite{PhysRevE.100.012137}. We also found that the entanglement entropy can be expressed as a series in particle number cumulants, with coefficients that agree with the well known result for free fermions in Gaussian states~\cite{song2012bipartite}.
    
    Next, extending this framework to an annular subsystem reveals that both R\'enyi entropies and cumulants decompose additively into individual contributions from the two boundaries of the annulus.  Numerical verification confirms that this additivity holds robustly, breaking down only in the limit where the  thickness of the annulus is very small.

    Looking forward, our analytical method extends well beyond the $l=0$ sector. The underlying diagonal form of the overlap matrix for radially symmetric subsystems remains robust for more general states with non-zero $l$, as long as we choose  the single particle states having distinct values of angular momenta, $\ell=(k-l)$. For such states, the present framework can be directly applied to compute the entanglement and FCS of higher excited states. Investigating how these configurations influence the behaviour of entanglement and FCS would be an interesting future direction.  
    
    \begin{acknowledgments}
        We acknowledge support of the Department of Atomic Energy, Government of India, under project no. RTI4001. AK acknowledges the financial support under project ANRF/ARGM/2025/001207/MTR from the ANRF, DST, Government of India. SNM acknowledges support from ANR Grant No. ANR-23-CE30-0020-01 EDIPS. A.D. acknowledges support from the J.C.~Bose Fellowship (JCB/2022/000014) of the Science and Engineering Research Board, Department of Science and Technology, Government of India. We thank the ICTS program, Indo-French workshop on Classical and quantum dynamics in out of equilibrium systems  (ICTS/ifwcqm2024/12), during which this collaboration was initiated.
        
    \end{acknowledgments}
    \newpage
    \begin{widetext}
        \appendix
        \section{Entanglement Entropy for the Disc Region}\label{appendix:ee_for_disc}
        \noindent
        We evaluate the integral $\mathcal{I}_{q}$ from Eq.~\eqref{eq:exact_integral_0}
        in the three regimes of the $N$-particle excited state. Specifically, we
        investigate its behaviour in the \textit{bulk} and \textit{edges} of the
        annular region occupied by the particles.
    
        \subsection{EE in the bulk}
        \label{sec:bulk_renyi}
        \noindent
        Noting that the radial coordinate $r=\sqrt{N}\,\xi$, with $\xi\sim O(1)$, the integration
        limits of $\mathcal{I}_{q}$ transform to
        $\frac{\sqrt{N}}{2\xi}(\mu-\xi^{2})$ and
        $\frac{\sqrt{N}}{2\xi}(1+\mu-\xi^{2})$. The integrand in Eq.~\eqref{eq:exact_integral_0} is peaked at $z=0$ and decays as $e^{-z^{2}}$. Hence, the dominant contribution to the integral comes from a region of $O(1)$ around $z=0$. In that region, the denominator of the argument of ${\rm erfc}$ function can be neglected. Hence, the function $\mathcal{I}_{q}$ takes the simpler form,
        \begin{align}
            \mathcal{I}_{q}(\mu,N; \sqrt{N}\xi) & \approx \frac{1}{\pi (1-q)}\int_{\frac{\sqrt{N}}{2\xi}(\mu-\xi^{2})}^{\frac{\sqrt{N}}{2\xi}(1+\mu-\xi^{2})}\,\diff{z}\,\ln\left[\left\{\frac{1}{2}\mathrm{erfc}(z\sqrt{2})\right\}^{q}+\left\{\frac{1}{2}\mathrm{erfc}(-z\sqrt{2})\right\}^{q}\right]. \label{eq:approx_integral}
        \end{align}
        In the domain $\mu<\xi^{2}<1+\mu$ and in the large-$N$ limit, this
        expression simplifies to
        \begin{align}
            \mathcal{I}_{q}(\mu,\infty;\sqrt{N}\xi)\approx [\sqrt{2}\pi(1-q)]^{-1}\int_{-\infty}^{\infty}\diff{z}\,\ln\left[\left\{\frac{1}{2}\mathrm{erfc}(z)\right\}^{q}+\left\{\frac{1}{2}\mathrm{erfc}(-z)\right\}^{q}\right]:=\sigma_{q}. \label{eq:approx_integral_1}
        \end{align}
        We observe that Eq.~\eqref{eq:approx_integral_1} is independent of both
        $\mu$ and $N$ in this range of $\xi$.

        \subsection{EE at the inner and outer edges}
        \label{sec:inner_edge_renyi}
        \noindent
        For large $N$, we now look at the neighbourhood of the inner edge defined
        by $\xi = \sqrt{\mu}+\frac{y}{\sqrt{N}}$, where $y=\mathcal{O}(1)$. Then
        we have
        \begin{align}
            \label{eq:change_variable}
            \begin{split}
            \frac{\sqrt{N}}{2\xi}(\xi^{2}-\mu)&\approx y, \\
            \frac{\sqrt{N}}{2\xi}(1+\mu-\xi^{2})&\approx \sqrt{\frac{N}{4\mu}}+\mathcal{O}(1).\end{split}
        \end{align}
        Under the assumption of finite $\mu$ and large $N$, the expression
        \eqref{eq:approx_integral} reduces to,
        \begin{align}
            \mathcal{I}_{q}\left(\mu,\infty;\sqrt{N \mu}+y\right)\approx [\sqrt{2}\pi(1-q)]^{-1}\int_{-\infty}^{y\sqrt{2}}\,\diff{z}\,\ln\left[\left\{\frac{1}{2}\mathrm{erfc}\left(z\right)\right\}^{q}+\left\{\frac{1}{2}\mathrm{erfc}\left(-z\right)\right\}^{q}\right]:= F_{q}(y). \label{eq:approx_integral_2}
        \end{align}
        which defines the function $F_{q}(y)$, that is independent of $\mu$ and
        $N$. At the outer edge, setting $\xi=\sqrt{1+\mu}+\frac{y}{\sqrt{N}}$, and
        performing a similar calculation gives:
        \begin{equation}
            \mathcal{I}_{q}\left(\mu,\infty;\sqrt{N(1+\mu)}+y\right) = F_{q}(-y),
            \label{eq:approx_integral_3}
        \end{equation}
        Eqs.~\eqref{eq:approx_integral_2} and \eqref{eq:approx_integral_3} show that
        the Rényi entropy exhibits mirrored behaviour at the inner and outer edges,
        differing only by a scaling factor of $2\pi r$, i.e. the circumference
        of the disc region under consideration.

        \section{Cumulants from CCGF}
        \label{app:ccgf_disc_appe}
        \noindent
        Following the approach in \cite{PhysRevA.99.021602}, we derive the full hierarchy of cumulants by Taylor-expanding the centered cumulant generating function (CCGF), $\chi(\lambda; r)$, in its sum form as given in Eq.~\eqref{eq:ccgf_sum},
        \begin{equation}
            \chi(\lambda;r)=\sum_{s=1}^{N}\left\{\ln\left[1+(\mathrm{e}^{-\lambda}-1)\,a_{s}(r)\right]+\lambda\, a_{s}(r)\right\}.
            \label{eq:ccgf_appendix_start}
        \end{equation}
        We first consider the regime $0<a_{s}(r)<\frac{1}{2}$. Factoring out $(1-a_{s}(r))$ from the logarithmic argument,
        \begin{equation}
            1+(\mathrm{e}^{-\lambda}-1)\,a_{s}(r)=\left(1-a_{s}(r)\right)\left[1+\mathrm{e}^{-\lambda}\,\frac{a_{s}(r)}{1-a_{s}(r)}\right].
        \end{equation}
        Eq.~\eqref{eq:ccgf_appendix_start} becomes
        \begin{equation}
            \chi(\lambda;r)=\sum_{s=1}^{N}\Big[\ln\left(1-a_{s}(r)\right)+\lambda\, a_{s}(r)+\ln\!\left(1+\mathrm{e}^{-\lambda}\,\frac{a_{s}(r)}{1-a_{s}(r)}\right)\Big].
            \label{eq:ccgf_factored_1}
        \end{equation}
        Since $0<a_{s}(r)<\frac{1}{2}$ implies $0<\frac{a_{s}(r)}{1-a_{s}(r)}<1$, we expand the logarithm into the power series $\ln(1+u)=-\sum_{l=1}^{\infty}(-u)^{l}/l$ with $u=\mathrm{e}^{-\lambda}\,\frac{a_{s}(r)}{1-a_{s}(r)}$, and subsequently expand each exponential as $\mathrm{e}^{-l\lambda}=\sum_{p=0}^{\infty}\frac{(-\lambda)^{p}}{p!}\,l^{p}$. Exchanging the order of the two sums gives,
        \begin{equation}
            \ln\!\left(1+\mathrm{e}^{-\lambda}\,\frac{a_{s}(r)}{1-a_{s}(r)}\right)=\sum_{p=0}^{\infty}\frac{(-\lambda)^{p}}{p!}\times\left\{-\sum_{l=1}^{\infty}l^{p-1}\left(-\frac{a_{s}(r)}{1-a_{s}(r)}\right)^{l}\right\}.
            \label{eq:double_series_1}
        \end{equation}
        The $p=0$ and $p=1$ terms of Eq.~\eqref{eq:double_series_1} can be resummed in closed form,
        \begin{align}
            -\sum_{l=1}^{\infty}\frac{1}{l}\left(-\frac{a_{s}(r)}{1-a_{s}(r)}\right)^{l}&=\ln\left(1+\frac{a_{s}(r)}{1-a_{s}(r)}\right)\nonumber\\
            &=-\ln\left(1-a_{s}(r)\right),\\
            (-\lambda)\left\{-\sum_{l=1}^{\infty}\left(-\frac{a_{s}(r)}{1-a_{s}(r)}\right)^{l}\right\}&=-\lambda\, a_{s}(r),
        \end{align}
        and exactly cancel the remaining logarithmic and linear contributions, $\ln\left(1-a_{s}(r)\right)$ and $\lambda a_{s}(r)$, in Eq.~\eqref{eq:ccgf_factored_1}. We are therefore left with
        \begin{equation}
        \chi(\lambda;r)=\sum_{p=2}^{\infty}\frac{(-\lambda)^{p}}{p!}\left\{\sum_{s=1}^{N}\sum_{l=1}^{\infty}\frac{(-1)}{l^{1-p}}\left(-\frac{a_{s}(r)}{1-a_{s}(r)}\right)^{l}\right\}.
        \end{equation}
        The inner sum over $l$ is precisely the defining series of the polylogarithm, $\mathrm{Li}_{s}(x)=\sum_{k=1}^{\infty}k^{-s}x^{k}$. Matching term-by-term with the standard expansion of the CCGF, $\chi(\lambda;r)=\sum_{p=2}^{\infty}\frac{(-\lambda)^{p}}{p!}\,\langle \hat{N}^{p}_{r}\rangle_{c}$, directly identifies the cumulants:
        \begin{equation}
        \langle \hat{N}^{p}_{r}\rangle_{c} = -\sum_{s=1}^{N}\mathrm{Li}_{1-p}\left(-\frac{a_{s}(r)}{1-a_{s}(r)}\right).
        \end{equation}
        Because the polylogarithm series converges only for $|x|<1$, i.e.\ $\frac{a_{s}(r)}{1-a_{s}(r)}<1$, this specific expression holds only when $0 \le a_{s}(r) < \frac{1}{2}$. To evaluate the regime where $\frac{1}{2} \le a_{s}(r) \le 1$, we instead factor out $\mathrm{e}^{-\lambda}a_{s}(r)$ from the logarithmic argument,
        \begin{equation}
            1+(\mathrm{e}^{-\lambda}-1)\,a_{s}(r)=\mathrm{e}^{-\lambda}a_{s}(r)\left[1+\mathrm{e}^{\lambda}\,\frac{1-a_{s}(r)}{a_{s}(r)}\right],
        \end{equation}
        so that Eq.~\eqref{eq:ccgf_appendix_start} now reads
        \begin{equation}
            \chi(\lambda;r)=\sum_{s=1}^{N}\Big[\ln a_{s}(r)-\lambda\left(1- a_{s}(r)\right)+\ln\!\left(1+\mathrm{e}^{\lambda}\,\frac{1-a_{s}(r)}{a_{s}(r)}\right)\Big].
            \label{eq:ccgf_factored_2}
        \end{equation}
        In this regime, $0<\frac{1-a_{s}(r)}{a_{s}(r)}<1$, and the same two-step expansion as before, now with $\mathrm{e}^{l\lambda}=\sum_{p=0}^{\infty}\frac{(-\lambda)^{p}}{p!}\,(-l)^{p}$, gives, upon noting that $-\frac{1-a_{s}(r)}{a_{s}(r)}=1-\frac{1}{a_{s}(r)}$,
        \begin{equation}
            \ln\!\left(1+\mathrm{e}^{\lambda}\,\frac{1-a_{s}(r)}{a_{s}(r)}\right)=\sum_{p=0}^{\infty}\frac{(-\lambda)^{p}}{p!}\,(-1)^{p}\times\left\{-\sum_{l=1}^{\infty}l^{p-1}\left(1-\frac{1}{a_{s}(r)}\right)^{l}\right\}.
            \label{eq:double_series_2}
        \end{equation}
        As before, the $p=0$ term of Eq.~\eqref{eq:double_series_2} simplifies to $\ln\left(1+\frac{1-a_{s}(r)}{a_{s}(r)}\right)=-\ln a_{s}(r)$, and the $p=1$ term to $\lambda\left(1-a_{s}(r)\right)$, exactly cancelling the remaining terms of Eq.~\eqref{eq:ccgf_factored_2}. Identifying the sum over $l$ with the polylogarithm series, we obtain
        \begin{equation}
        \chi(\lambda;r) = \sum_{p=2}^{\infty}\frac{(-\lambda)^{p}}{p!}\left\{-\sum_{s=1}^{N}\,(-1)^{p}\mathrm{Li}_{1-p}\left(1-\frac{1}{a_{s}(r)}\right)\right\}.
        \end{equation}
        Combining both domains allows us to evaluate the exact cumulants across the entire spectrum of the overlap matrix eigenvalues, $a_s(r)$:
        \begin{equation}
            \langle \hat{N}^{p}_{r}\rangle_{c}=\sum_{s=1}^{N}\, n_{p}(a_{s}(r)),
        \end{equation}
        where:
        \begin{equation}
            n_{p}(x)=\begin{cases}
                -\mathrm{Li}_{1-p}\left(-\frac{x}{1-x}\right), & 0<x<\frac{1}{2} \\ 
                (-1)^{p-1}\mathrm{Li}_{1-p}\left(1-\frac{1}{x}\right), & \frac{1}{2}<x<1.
            \end{cases}
        \end{equation}
        The cumulants for the annular region can be calculated in the same manner by replacing the corresponding overlap matrix eigenvalues.
        \section{Cumulants at the large N limit}
        \label{appendix:particle_num_cum_approx}
        \noindent
        We evaluate the integral $\mathcal{J}_{N}^{p}$ in Eq.~\eqref{eq:cumulant_integral_0}
        in a similar way to the Appendix~\ref{appendix:ee_for_disc}.
        \subsection{Cumulants in the bulk}
        \label{sec:cumulants_in_bulk}
        \noindent
        We can use the changes of variable and the limits of integration used
        for the bulk R\'enyi entropy case in the appendix~\ref{sec:bulk_renyi}
        and write the function $\mathcal{J}^{p}_{N}$ as,
        \begin{align}
            y(k;\sqrt{N}\xi)                      & \approx z\sqrt{2}+\mathcal{O}(N^{-1/2}),\nonumber\\
            \mathcal{J}^{p}_{N}(\mu; \sqrt{N}\xi) & \approx \frac{1}{\pi}\, \int_{\frac{\sqrt{N}}{2\xi}(\mu-\xi^{2})}^{\frac{\sqrt{N}}{2\xi}(1+\mu-\xi^{2})}\,\diff{z}\,n_{p}\left(\tfrac{1}{2}\mathrm{erfc}(z\sqrt{2})\right).\label{eq:approx_j_integral}
        \end{align}
        In the domain $\mu<\xi^{2}<1+\mu$ and the limit of large $N$, the
        expression simplifies to
        \begin{equation}
            \mathcal{J}^{p}_{\infty}(\mu;\sqrt{N}\xi)\approx [\sqrt{2}\pi]^{-1}\int
            _{-\infty}^{\infty}\diff{z}\,n_{p}\left(\tfrac{1}{2}\mathrm{erfc}\left(z\right)\right):=\kappa_{p}.\label{eq:approx_j__integral_1}
        \end{equation}
        In particular, for odd $p$, the integrand in Eq.~\eqref{eq:approx_j__integral_1}
        is an odd function. Hence, $\kappa_{p}$ vanishes for odd $p$.
        \subsection{Cumulants at the inner and outer edges}
        \label{sec:cumulants_edge}
        \noindent
        In the neighbourhood of the inner edge defined by $\xi= \sqrt{\mu}+\frac{y}{\sqrt{N}}$,
        the limits of integral \eqref{eq:approx_j_integral} change as in the section
        \ref{sec:inner_edge_renyi},
        \begin{equation}
            \mathcal{J}^{p}_{\infty}(\mu;\sqrt{N\mu}+y)\approx [\sqrt{2}\pi]^{-1}
            \int_{-\infty}^{y\sqrt{2}}\diff{z}\,n_{p}\left(\tfrac{1}{2}\mathrm{erfc}(-z)\right):=j_{p}(y),\label{eq:cumul_integ_inner}
        \end{equation}
        defining $j_{p}(y)$ as a function independent of $\mu$ and $N$. An analogous
        calculation at the outer edge $\xi=\sqrt{1+\mu}+\frac{y}{\sqrt{N}}$ gives:
        \begin{equation}
            \mathcal{J}_{\infty}^{p}\left(\mu,\infty;\sqrt{N(1+\mu)}+y\right)=j_{p}
            (-y).\label{eq:cumul_integ_outer}
        \end{equation}
    \end{widetext}
    \bibliography{refs}
\end{document}